\documentstyle[prb,aps,psfig]{revtex}   
\begin{document} 
\title{Landau level mixing and spin degeneracy in the quantum Hall effect}
\author{V. Kagalovsky*, B. Horovitz and Y.Avishai}
\address{Department of Physics, Ben-Gurion University of the Negev, 
Beer-Sheva, Israel} 
\maketitle
\begin{abstract}
We study dynamics of electrons in a magnetic field using 
a network model with two channels per link with random mixing in a
random intrachannel potential; the channels represent either 
two Landau levels or two spin states. We consider channel mixing 
as function of the energy separation of the two extended states and show 
that its effect changes from repulsion to attraction as the energy separation 
increases. For two Landau levels this leads to level floating at low 
magnetic fields while for Zeeman split spin states we predict level 
attraction at high magnetic fields, accounting for ESR data. We also study  
random mixing of two degenerate channels, while the intrachannel potential is 
periodic (non-random). 
We find a single extended state 
with a localization exponent $\nu\approx 1.1$ for real scattering at nodes; 
the general case has also a single extended state, though the
localized nature of nearby states sets in at unusually large scales.
\end{abstract}


\section{Introduction}
\label{intro}
The Quantum Hall Effect (QHE) remains
 a great attraction for theoreticians and 
experimentalists. Of particular interest is the divergence of the 
localization length at a discrete set of energies, corresponding to 
extended states. This has motivated a large variety of methods for 
studying the pertinent metal insulator transition as a critical 
second order phase transition \cite{rqhe}, such as field theoretical methods 
\cite{rpruis}, semiclassical methods \cite{rtrugman,rmyl}, numerical 
methods \cite{rAndo} and finite size scaling for transfer matrices 
\cite{rchalk}.
Most of these works focus on properties near an isolated extended state 
with emphasis on the critical exponent $\nu$ of the localization length. 
The most recent  
experimental result \cite{rkoch1} $\nu =2.3\pm 0.1$ is in 
agreement with theoretical predictions \cite{rmyl,rchalk}.

The situation with a few extended states, allowing for coupling 
between these states received less attention. This situation is relevant to 
the behavior of
delocalized  states in weak magnetic fields. It is well known
that the existence of delocalized states is a necessary 
condition for the QHE behavior. On the other hand, scaling theory 
\cite{randers} and numerical studies \cite{rnumer}
imply that in a 2D system in the absence of a magnetic field all states 
should be localized. Therefore , a scenario in which 
delocalized states "float up" above the Fermi level as the  
magnetic field decreases has been suggested \cite{rkhmel,rlaugh}. 
Recent experiments 
\cite{rexp}  
show that, indeed, the energy of 
of the lowest delocalized state floats up above the Fermi level as 
magnetic field is reduced. This corresponds to a transition from an 
insulator to a quantum Hall conductor at both low and high fields. 
Numerical 
studies of a few Landau bands\cite{rlee,rleeko,rwang,rhanna,rjap} focused on 
critical exponents except for an early work by Ando \cite{rAndo} which 
supports the floating scenario. Ando's work used $\delta$ function 
impurities which is, however, not suitable when the impurity 
concentration is too low since bound states of the $\delta$ potential 
shift the extended state even for a single Landau band \cite{rAndo}.
We have recently shown \cite{rour1} that two extended states attract
 each other which leads to a minimum in the energy of the lower state. 
This attraction has been also predicted theoretically 
by Shahbazyan and Raikh 
\cite{rraih} using a  
high magnetic field expansion.

The system of two coupled extended states was also studied in the context 
of a spin-split Landau band \cite{rlee,rleeko,rwang,rhanna,rjap}. 
In the absence of a 
Zeeman term it was found that two separate extended states appear, each 
with a localization exponent of $\approx 2.3$. Allowing for a finite 
Zeeman term leads to level attraction at high fields \cite{rour1}. In 
fact, electron spin resonance (ESR) data \cite{dobers} has shown that 
the spin splitting has an unusual nonlinear dependence on field. As 
shown below, this nonlinear dependence is consistent with 
localization phenomena and level attraction of extended states.

A different type of system where two degenerate states in a strong 
field (e.g. spin states in the first Landau level) are coupled by 
random mixing, but the scalar (i.e. intrachannel) potential is absent,
 was recently studied \cite{rhanna,rjap,rhikami,dkkl}.
This corresponds to a spin-orbit coupling which is dominant relative
 to the scalar potential scattering. 
For white noise inter-band mixing \cite{rhikami} it was found that an 
extended state at the original Landau level energy exists, suggesting a 
new universality class. For smooth inter-band disorder \cite{dkkl}, it was
suggested that there are two separate extended states as in the 
spin-split case in addition to a third extended state in between, at 
the original Landau level.

In the present work we study these and other aspects, using 
various extensions of the network 
model introduced by Chalker and 
Coddington (CC) \cite{rchalk}. In the network model 
electrons move along unidirectional links which form a closed 
loop in analogy with 
semiclassical motion on contours of constant potential. Scattering 
between links is allowed at nodes, in analogy with tunneling through 
saddle point potentials in the semiclassical model. The assumption that 
each link carries current only in one direction implies that the 
wavepackets are sufficiently localized in the transverse direction, i.e. 
the magnetic length is small in comparison with the spacing of nodes or 
with the correlation length of the potential fluctuations.

Our paper is organized as follows. In Sec.\ \ref{describe} we describe 
the network model for one and two channels and relate parameters 
of the system to the transfer matrix. 
In Sec.\ \ref{u2} , we expand our earlier work \cite{rour1} and 
study the two-channel network model corresponding
to two coupled Landau bands. 
We evaluate the energies of extended states as function of $\Delta$, the 
bare energy separation of extended states in the absence of level mixing.
In Sec.\ \ref{degen} we investigate the two-level system 
with random mixing while the scalar potential scattering is 
periodic (i.e. non-random) and consider different types of
symmetries for random mixing. Our results 
are summarized in Sec.\ \ref{concl} where the ESR data is also 
discussed. In Appendix A, 
using the one-channel network model, we test a possibility of 
sublocalization behavior of the wavefunction, based on a 
possible fractal behavior of the nodes. In Appendix B the tunneling 
amplitude between Landau levels is estimated.

\section {The Network Model}
\label{describe}

Consider first a one-channel CC network 
\cite{rchalk}, which has directed links and scattering at nodes 
(Fig.\ \ref{picture}a).
Propagation along links yields a random phase 
$\phi$, thus links are
presented by diagonal matrices with diagonal elements in the form
$\exp (i\phi)$. Transfer 
matrix for one node relates a pair of incoming and outgoing links 
on the left (links $2,4$ in Fig.\ \ref{picture}b) to a pair 
of links on the right (links $1,3$); it has the form 
\begin{equation}
{\bf T}=\begin{array}{ccc}
\left( \begin{array}{cc} e^{i\alpha_1} & 0  \\
 0 & e^{i\alpha_2}
\end{array}
\right) &
\left( \begin{array}{cc}
\cosh\theta & \sinh\theta  \\
 \sinh\theta & \cosh\theta
\end{array}
\right) &
\left( \begin{array}{cc} e^{i\alpha_3} & 0  \\
 0 & e^{i\alpha_4}
\end{array}
\right)
\end{array}
\label{first}
\end{equation}

In order for the system to be invariant, on average, under $90^{\circ}$ 
rotation
the next neighbor node is obtained by rotating Fig.\ \ref{picture}b by 
$90^{\circ}$ and writing the states on links $4,1$ in terms of those 
on links $2,3$. 
The transfer matrix has then the same form as 
in Eq.\ (\ref{first}) with a 
parameter $\theta '$ replacing $\theta$ where \cite{rchalk} $\sinh\theta '
=1/\sinh\theta$  
and the phases $\alpha_1,\alpha_2,\alpha_3,\alpha_4$ are replaced by 
$-\alpha_3,\alpha_1,\alpha_4+\pi,-\alpha_2$ (Fig.\ \ref{picture}c). 
We therefore describe scattering at the nodes 
indicated in Fig.\ \ref{picture} by circles with transfer matrix ${\bf
  T}(\theta)$ and 
at the nodes 
indicated by boxes with ${\bf T}(\theta ')$. 

The phases $\alpha_i$ can be absorbed into phases describing propagation 
on links. Since a shift by a common phase
 in all links of a given column does not 
affect the Lyapunov exponents of the network, we can choose the 
phases on the links as $\pm\alpha, \pm\beta$ (Fig.\ \ref{picture}a) 
where $\alpha =\frac{1}{2}\alpha_1-\frac{1}{2}\alpha_4-\alpha_2$, $\beta =
\frac{1}{2}\alpha_1-\frac{1}{2}\alpha_4+\alpha_3$. Note that the sum of left 
links equals those of right links $+\pi$. For a random potential 
(Section\ \ref{u2}) these link phases are considered as random while for 
a periodic potential (Section\ \ref{degen}) we choose specific sets. The
appropriate procedure 
is then to multiply transfer matrices for links and nodes 
alternately and derive Lyapunov exponents for strips of width up to 
$64$ and of length 
of typically $60,000$ units (Section III) or $240,000$ (Section
IV). At these lengths our error for the localization length (i. e. for
the inverse of the smallest Lyapunov exponent) is $\lesssim 0.5\%$.

In the following we relate  the node parameter $\theta$ to the electron
energy by using known results for scattering from a saddle point 
potential\cite{rfert}. 
The transmission 
probability $T$ of an electron with energy $E$ through a saddle-point 
potential $V_{SP}(x,y)=-U_xx^2+U_yy^2+V_0$ in a perpendicular magnetic 
field $B$ is given by  
\begin{equation}
T=\frac{1}{1+\exp (-\pi\epsilon )},
\label{sixth}
\end{equation}
where 
\begin{equation}
\epsilon\equiv (E-(n+\frac{1}{2})E_2-V_0)/E_1,
\label{eqeps}
\end{equation}
 and 
\begin{equation}
E_1=\left[\Omega\left\{\gamma^2+\left(\frac{\omega_c}{4}\right)^2
\right\}^{1/2}-\frac{1}{4}\Omega^2-\left(\frac{\omega_c}{4}
\right)^2\right]^{1/2},
\label{seventh}
\end{equation} 
where $\Omega =(\frac{1}{4}\omega_{c}^2+(U_y-U_x)/m)^{1/2}$, $\gamma =
(U_y+U_x)/4m\Omega$, $m$ is the electron mass and $\omega_c=eB/mc$ is 
the cyclotron frequency. The oscillator frequency $E_2$ is
\begin{equation}
E_2=2\left[\Omega\left\{\gamma^2+\left(\frac{\omega_c}{4}\right)^2
\right\}^{1/2}
+\frac{1}{4}\Omega^2+\left(\frac{\omega_c}{4}\right)^2\right]^{1/2}.
\label{eighth}
\end{equation}

From Eq.\ (\ref{sixth}) the ratio of reflection and transmission coefficients 
is $\exp (-\pi\epsilon)$. On the other hand 
the transfer matrix Eq.\ (\ref{first}) determines this ratio as 
$(\sinh\theta )^2$. 
Therefore, the relation between node parameter $\theta$ in CC model 
and the electron energy is
\begin{equation}
\epsilon =-\frac{2}{\pi}\ln(\sinh\theta )
\label{ninth}
\end{equation}

We note that unlike discrete Landau levels, the 
saddle point potential 
allows for a continuous energy E for each discrete state $n$. Furthermore, 
for $\omega_c \gg \gamma\ \  E_1 \rightarrow 0$, 
$E_2 \rightarrow \omega_c$ and $E \rightarrow \hbar 
\omega_c (n+1/2) + V_0$ corresponds to discrete Landau levels. In the 
opposite limit of $\omega_c \ll \gamma$ we have $E_1\rightarrow 
(U_x/2m)^{1/2}$, $E_{2}\rightarrow (2U_{y}/m)^{1/2}$ and the integer $n$ 
corresponds  to a 
quantum number of the harmonic potential $U_y y^2$. 

In the case of one channel
per link a single extended state \cite{rchalk} 
is at $\epsilon =0$ (i.e. $\theta =0.8814$), 
corresponding to the center of any band $n$, $E=(n+1/2)E_2+V_0$.
Numerical studies of this 
system with width $M$ and periodic boundary conditions confirm
the one parameter scaling hypothesis, i.e. the localization length $\xi_M$ 
is given by a scaling function $f$, where
\begin{equation}
\frac{\xi_M}{M} =f\left(\frac{M}{\xi (E)}\right)
\label{second}
\end{equation}
with $\xi (E)\sim |E|^{-\nu}$ and $\nu=2.5\pm 0.5$. This result is in 
 good agreement with experimental data for spin resolved levels 
\cite{rkoch1}, numerical simulations using other models \cite{rhanna,rhuck1}
and semiclassical derivation \cite{rmyl,rour} that predicts $\nu =7/3$. 
We have repeated the one-channel calculation (Appendix A) with a particular 
emphasis on the possibility of sublocalization; we find that localized 
states decay as a regular exponential.

A two channel network, i.e. two channels per link, is  
characterized by parameters  $\theta$ and $\theta -\Delta\theta$ 
which determine the tunneling amplitude at the node for each channel with 
$\Delta\theta$ related to the relative energy interval between the
bands of the two channels.
The transfer matrix is 
parameterized in the following way \cite{rleeko}: 
\begin{equation}
{\bf T}=\begin{array}{ccc}
\left( \begin{array}{cc}{\bf U_1} & 0  \\
 0 & {\bf U_2}
\end{array}
\right) &
\left( \begin{array}{cc}{\bf C} & {\bf S}  \\
 {\bf S} & {\bf C}
\end{array}
\right) &
\left( \begin{array}{cc}{\bf U_3} & 0  \\
 0 & {\bf U_4}
\end{array}
\right) .
\end{array}
\label{tmatr}
\end{equation}
The transfer matrix at a node is composed of the blocks:
\begin{equation}
{\bf C}=\left( \begin{array}{cc}
\cosh\theta & 0 \\
0 & \cosh (\theta -\Delta\theta )\end{array}
\right),\ \
{\bf S}=\left( \begin{array}{cc}
\sinh\theta & 0
\\
0 & \sinh (\theta -\Delta\theta )
\end{array}
\right).
\label{nodematr0}
\end{equation}
Defining an 
energy parameter $\Delta$ such that $\epsilon -\Delta =-(2/\pi )
\ln (\sinh (\theta -\Delta\theta )$ we obtain 
\begin{equation}
{\bf C}=\left( \begin{array}{cc}
\sqrt{1+\exp [-\pi(\epsilon-\Delta)]} & 0 \\
0 & \sqrt{1+\exp [-\pi\epsilon]} \end{array}
\right),\ \
{\bf S}=\left( \begin{array}{cc}
\exp [-\pi(\epsilon-\Delta)/2] & 0
\\
0 & \exp [-\pi\epsilon/2]
\end{array}
\right) .
\label{nodematr}
\end{equation}
Propagation  along links yields random phases $\phi_i ,i=1-4$ and also
allows mixing between two different channels. It 
is described by block ${\bf U}$: 
\begin{equation}
{\bf U}=\begin{array}{ccc}
\left( \begin{array}{cc}e^{i\phi_1} & 0  \\
 0 & e^{i\phi_2}
\end{array}
\right) & 
\left( \begin{array}{cc}\sqrt{1-x^2} & -x  \\
 x & \sqrt{1-x^2}
\end{array}
\right) &
\left( \begin{array}{cc}e^{i\phi_3} & 0  \\
 0 & e^{i\phi_4}
\end{array}
\right) ,
\end{array}
\label{umatr}
\end{equation}
where $x^2$ is the mixing probability between different levels. 

We note that the maximal number of independent parameters in 
a $U(2)$ matrix presented in Eq. (\ref{umatr}) is four. 
These phases can be chosen as $\delta=1/2(\phi_1+\phi_2+
\phi_3+\phi_4), 
\delta_1=-1/2(\phi_2+\phi_4), \delta_2=-1/2(\phi_2-\phi_3)$ 
so that
\begin{equation}
{\bf U}=e^{i\delta}\left( \begin{array}{cc}
e^{i\delta_1}\sqrt{1-x^2} & 
-e^{i\delta_2}x \\
e^{-i\delta_2}x & e^{-i\delta_1}\sqrt{1-x^2} 
\end{array} \right) .
\label{umatr1}
\end{equation}
The phase $\delta$ 
corresponds to a scalar potential, it can be either random (Section\ \ref
{u2}) or fixed for a periodic  potential (Section\ \ref{degen}). 
If $\delta =
2\pi\times$integer the ${\bf U}$ matrix changes its symmetry group from 
$U(2)$ to a unitary unimodular $SU(2)$. 
In Section\ \ref{degen} we study the effect of various subgroups of $U(2)$ 
on the critical properties of the system.

\section{Non-degenerate levels with random potential and random mixing}
\label{u2}

We study here a system of two Landau levels in the presence of a smooth 
random potential, so that for proper description we use the most 
general form of Eq. (\ref{umatr1}) with four random variables. 

In the absence of level mixing we know from results of the one channel 
system \cite{rchalk} that $\epsilon = 0$ corresponds to extended states. 
 This defines "bare" extended states at $E_{ex} 
=E_2 (n+1/2) + V_0$ and the bare energy splitting of the n=0,1 states is 
then 
$E_2$. Note that the splitting $E_2$ which is 
magnetic field dominated at $\omega_c(m/U)^{1/2}>1$ remains finite 
as $\omega_c\rightarrow 0$ and is 
potential dominated at $\omega_c(m/U)^{1/2}<1$ (here $U_x=U_y\equiv U$). The 
latter region is acceptable for a network model if 
the correlation length of the potential fluctuation is long compared with 
the magnetic length, so that locally the saddle point potential 
determines a finite splitting. The splitting $\Delta =E_2/E_1$ is 
therefore bounded by $\Delta\ge 2$ at $\omega_c\rightarrow 0$.
 
The application to $n=0,1$ Landau levels assumes that mixing of states 
with $n$'s differing by $\Delta n=2$ is much smaller than for those 
with $\Delta n=1$. Mixing or transition rates can be evaluated for a 
potential of the form $(1/2)\tilde{U}(x^2+y^2)-\tilde{U}y^3/\lambda$ 
where $\lambda$ is a measure of the correlation length of the random 
potential. For states near the local minimum we find that mixing is 
given by Eq.\ (\ref{last}) (see Appendix B).
We therefore expect 
our results for the $n=0$ state, within the two channel model, to be 
valid down to $m\omega_{c}^{2}/U\approx 1/\ln (\lambda /\ell )\lesssim 1$ 
(assuming $\tilde{U}
\approx U$) so that the range of valid $m\omega_{c}^{2}/U$ values 
is extended down to lower limits for longer range potential 
fluctuations. The results for the $n=1$ state are, however, not 
directly relevant.  

We consider below also the case of a spin split single
level $n=0$ for which $V_0=\pm (1/2)g^{*}\mu_B$, 
where $g^{*}$ is the electron $g$ factor and $\mu_B$ is the Bohr magneton. The
bare extended states correspond then to $E_{ex}=\frac{1}{2}E_2 \pm
\frac{1}{2}g^{*}\mu_B$ so that $\Delta = g^{*}\mu_B/ E_1$.

The system (see Eqs.\ (\ref{tmatr},\ref{nodematr})) is, on average, 
invariant under $90^{\circ}$ rotation 
if at the next neighbor node the transmission and reflection (of each 
channel) are interchanged, i.e. $\epsilon \rightarrow -\epsilon$ and 
$\epsilon -\Delta \rightarrow -\epsilon +\Delta$. The system is then 
symmetric under $\epsilon \rightarrow -\epsilon$, $\Delta \rightarrow 
-\Delta$ and the extended states at $\epsilon_i\ (i=1,2)$ become 
$\epsilon_i(-\Delta)=-\epsilon_i(\Delta)$. A further translation of 
energies by $\Delta$ returns the system to itself except for 
$1\leftrightarrow 2$ interchange, 
$\epsilon_2(-\Delta )+\Delta =\epsilon_1(\Delta)$, hence 
$\epsilon_1(\Delta)+\epsilon_2(\Delta)=\Delta$.
 Therefore results for the two energies $\epsilon_{1,2}$ are 
constrained by the condition $\epsilon_1 +\epsilon_2 =\Delta$.

We proceed further by  
calculating 
normalized localization length $\xi_M/M_l$ for systems with 
$M=16, 32$ (size of ${\bf T}$ matrix is $M\times M$ and $M_l=M/2$ is 
the number of links across the strip) and different values of 
$\epsilon$ and $\Delta$. For $\Delta =0$ we used also $M=64$ which 
affected critical energies by $3\%$, which are also within $5\%$ from 
the result of Wang et al. \cite{rwang}. 
The coupling $x$ is chosen to be uniformly distributed in the interval
$[0,1]$; 
we checked that other distributions in $x$ lead to similar results. 
 Finite size scaling is then used for fitting 
our data onto a single curve $f(\xi_{\infty} (\epsilon)/M_l)$ 
extracting values of localization length $\xi_{\infty}$ for the infinite 
2D system. 
Finally we look for the critical energies $\epsilon_i$ and critical 
exponent $\nu$ by requiring $\xi_{\infty}\sim|\epsilon 
-\epsilon_i|^{-\nu}$.
 The raw data for one particular $\Delta$ (see Fig.\ \ref{e32}) represents 
the  
characteristic features of the system. One can see that the values of 
$\xi_M/M_l$ for any two energy values whose sum equals $\Delta$ coincide as 
we expect from the symmetry condition. Another important feature is that for 
$\Delta\gtrsim 1$ we have two 
pronounced maxima of $\xi_M/M_l$, which we expect to be near the critical 
energies; this is supported by the scaling procedure. We find that 
approaching the critical energies from outside yields $\nu =2.5\pm 0.5$. 
The states between the two critical energies seem to be localized, i.e.
$\xi_M/M_l$ decreases with $M$, although the decrease is rather slow. 

The critical values 
$\epsilon_{1,2}$ are presented as a function of $\Delta$ in Fig.\ 
\ref{eps}. We  
cannot calculate $\xi_M/M$ for $\Delta >3.5$ because of round-off 
errors. The $\Delta\rightarrow\infty$ case can be solved analytically 
\cite{rraih} since then near one extended state the other channel is 
very far from tunneling and its trajectory is a closed loop between 
the nodes. Eliminating the closed loop variables and assuming that it 
mixes with links of only one node 
leads to extended states at 
$\epsilon_1(\Delta )\sim 1/\Delta$ and $\epsilon_2(\Delta )-
\Delta \sim -1/\Delta$. The rather flat behavior  of $\epsilon_1(\Delta )$ 
up to $\Delta <3.5$ implies that the asymptotic behavior 
\cite{rraih} sets in at a higher $\Delta$. 

The most remarkable aspect of the data is the crossing of 
$\epsilon_i(\Delta)$ with the bare extended states at $\Delta \approx 
0.5$. Usually one expects that mixing affects mainly extended states 
leading to level repulsion of these states. This expectation leads to 
$\epsilon_1(\Delta)$ approaching zero from below and $\epsilon_2(\Delta)$ 
approaching $\Delta$ from above, as implied by Fig. 2 of Ref. 
\onlinecite{rlee}. 
Contrary to this expectation we find that above $\Delta \approx 0.5$ 
there is level attraction. 

Level attraction is also supported by the following observation: 
Parameterize the upper branch of critical $\epsilon$ by $\theta$, i.e. 
$\epsilon =-(2/\pi )\ln\sinh\theta$; at $\Delta =0$ $\epsilon =0.23$, 
i.e. $\theta =0.65$ \cite{rlee,rwang}, and $\epsilon$ increases with 
$\Delta$ so that $\theta\rightarrow 0$. In the level repulsion 
scenario $\epsilon >\Delta$ so that the parameter $\Delta\theta$ 
in Eq.\ (\ref{nodematr0}) 
varies in the range $0\geq\Delta\theta\geq -0.88$. We could 
however define the model with parameters $\theta$, $\Delta\theta$ 
(instead of $\epsilon$, $\Delta$) with $\Delta\theta$ unbounded 
which implies that $\epsilon -\Delta$ must change sign.

 We present now results for the energies of extended states as a function 
of magnetic field by relating $\Delta$ to $B$ via Eqs. (5,6). As noted 
above $E_2$ is always finite so that $\Delta\ge 2$ ($\Delta \rightarrow 2$ 
for $B \rightarrow 0$ 
and $U_x=U_y$) so that we are always in the level attraction regime. The 
results are shown in the inset of Fig.\ \ref{eps} with the  
$\Diamond$ symbol (assuming for simplicity 
$U_x=U_y\equiv U$) and the full line is the lower bare extended state 
energy. Our data shows a 
minimum at $\omega_c (2m/U)^{1/2}\approx 0.5$ in the lower 
state, consistent with floating, and is a result of level attraction due to 
Landau level mixing. Allowing for mixing with the $n=2$ Landau level 
may cause floating of the $n=1$ state as well, but will have a small 
effect on the floating of the $n=0$ state, as discussed above. 
For N 
Landau levels our symmetry argument shows that the 
extended state energies come in pairs whose average is the same as for 
the bare states, i. e. $\epsilon_i +\epsilon_{N+1-i} =(N-1)\Delta$ with 
$i=1,\dots,N$. Hence we expect that the 
energies of the lower half states increase at low fields consistent with 
the floating scenario and with the experimental data \cite{rexp}.

The assumption of full mixing, i.e. $x\!\in\! [0,1]$ in Eq. (\ref{umatr})
 , is not valid 
for strong magnetic fields where tunneling between Landau levels should 
be suppressed. 
We model this situation in a reduced mixing model (see Appendix B), where 
the parameter $x$ in Eq. (12) is
now chosen randomly in the reduced range of $[0, \exp(-m\omega_c^2/U)]$. 
The results are shown in the inset
 of Fig.\ \ref{eps}, with the 
$\times$ symbol. The minimum in the lower level is now more pronounced 
and is at a higher field, $\omega_c (2m/U)^{1/2}\approx 1$.

Finally, we present application of our data to extended state energies for a
Zeeman spin splitting of a Landau band where $\Delta=g^*\mu_B B/E_1$. Since
$\Delta$ can cross the value $\approx 0.5$ we predict that spin splitting is
larger than expected for small fields (level repulsion) and is smaller than
expected for large fields (level attraction). In fact, we claim that level
attraction accounts for ESR data \cite{dobers} which shows a nonlinear
dependence of the spin splitting on B. In particular the data for $N=0$ in the
range $8-14$T is visibly nonlinear in the range $8-12$T.

To demonstrate the effect we consider $g^* \mu_B B/\omega_c=g^* m/2m_0=0.018$
(e.g. $g^*=0.51$ with an effective to free electron mass ratio of 
\cite{Lassing}
$m/m_0=0.07$) and replot the data of Fig.\ \ref{eps} for the splitting
$\Delta E=E_1(\epsilon _2-\epsilon_1)$ in Fig.\ \ref{diff}. 
The experimental data at $B>8$T fits
our results if $(U/m)^{1/2}=310$GHz is chosen (see section V for an independent
data leading to this value); the field scale is then $\omega_c
(U/m)^{1/2}\approx 1.3 B$(T). The deviation from a straight line measured at
$8-12$T cannot be accounted for by band structure calculations (see section V).
The data, therefore, provides a strong support for level attraction at these
fields.

\section{Degenerate Levels with random mixing}
\label{degen}

We consider in this Section a single Landau level with two degenerate 
spin states (i.e. no Zeeman term) where the only randomness comes from 
mixing of the spin states, e.g. by spin-orbit scattering. A model of this 
type was studied by Hikami, Shirai and Wegner \cite{rhikami} with the 
Hamiltonian
\begin{equation}
H=\frac{1}{2m}({\bf p}+{\bf A})^2+{\bf h}({\bf r}){\bf\cdot\sigma}.
\label{pauli}
\end{equation}
where ${\bf h}({\bf r})$ is a random field in $x,y$ directions and 
projection to states of the lowest Landau level is understood. This 
model has an extended state at the original Landau level \cite{rhikami} 
($\epsilon =0$) and possibly additional two extended states \cite
{dkkl,rhanna,rjap}, symmetric around $\epsilon =0$.

We wish to study the Hamiltonian (Eq.\ (\ref{pauli})) by a network 
model. We therefore replace the continuum by a periodic potential 
so that the nodes of the network are the periodic set of saddle points. 
In general, a magnetic flux through a unit cell leads to Aharonov-Bohm 
phases on the links which increase linearly from, say, left to right. This 
however reduces the symmetry around $\epsilon =0$ as checked by our 
simulations; this may be related to the formal loss of invariance 
under $90^{\circ}$ rotations. We therefore choose the unit cell to 
have an integer number of flux quanta. 

As shown in Figs. 1a-1b the phases on the links can be represented by
two phases $\alpha$, $\beta$. In order to choose relevant values of 
$\alpha$, $\beta$ we consider the pure system without spin mixing 
and try to make its spectra similar to that of the ${\bf p}^2/2m$ term in 
Eq. (13), i.e. a constant, with a total number of states close to that
of a Landau level. Extended states in the pure system can be found by
applying transfer matrices across a unit cell (i.e. two nearby nodes 
in Fig. 1) which by Bloch's theorem lead to multiplication by $\exp 
(iq)$ along the strip or $\exp (ik)$ in the direction across the
strip. This procedure yields the dispersion relation

\begin{equation}
\cosh (\pi\epsilon /2)=\frac{\sin q+\sin (k-\alpha +\beta )}{2\sin
  (\alpha
 +\beta )}.
\end{equation}

This relation shows that there is a maximal energy $\epsilon_{max}$ 
in the band where $\cosh (\pi\epsilon /2)=1/\sin (\alpha +\beta
)$. The density of states is linear at $\epsilon\rightarrow 0$ and 
saturates at $\epsilon_{max}$. The total number of states in the band
approaches that of a Landau level for $\alpha +\beta\rightarrow 0$. 
However, at $\alpha +\beta =0$ the spectra Eq. (14) is singular, so
that it seems that a small but finite value of $\alpha +\beta$ is needed.

The spin term in Eq.\ (\ref{pauli}) is taken to mix spin states on 
links. Since all links are unidirectional the transfer matrix is 
equivalent to an evolution operator which is a 
rotation in spin space $\exp [(i/\hbar )\int {\bf h}({\bf r})
{\bf\cdot\sigma}dt]$. Since successive rotations about the $x$,$y$
axes produce 
a rotation about the $z$ axis ${\bf h}$ should be a $3$D vector. 
The effect of these rotations is equivalent to an $SU(2)$ transfer
matrix (Eq.\ (\ref{umatr1}) with $\delta =0$) and 
we can interpret all independent phases as corresponding angles of 
the pseudofield. Below we consider such transfer matrices with different 
sets of independent parameters for the case of two degenerate levels. 

Consider first the case when ${\bf T}$ is a 
real matrix, i.e. the pseudofield is only in the $y$-direction and  
${\bf T}$ has then $SO(2)$ symmetry,
\begin{equation}
{\bf U}=\left( \begin{array}{cc}
\cos\phi & 
-\sin\phi \\
\sin\phi & \cos\phi 
\end{array} \right).
\label{uma}
\end{equation} 
The results for the case with the Haar measure,
i.e. $\phi\!\in\! [0,2\pi ]$ with $\cos\phi$ uniformly distributed, 
are shown in Fig.\ \ref{u1raw}. The data shows a single extended 
state at $\epsilon =0$. 

We know that a random $U(2)$ transfer matrices lead to two extended states. 
It is therefore interesting to modify the $SO(2)$ system by a random 
potential leading to a $U(1)\times SO(2)$ system   
\begin{equation}
{\bf U}=e^{i\delta}\left( \begin{array}{cc}
\cos\phi & 
-\sin\phi \\
\sin\phi & \cos\phi 
\end{array} \right).
\label{u1u1}
\end{equation} 
The data maintains 
a single peak at $\epsilon =0$ in this case as well.
Fitting the data of those two cases 
by a smooth function $f$ so that $\xi_M/M_l=
f(\epsilon M_{l}^{1/\nu})$ yields $\nu\approx 2.2$ for the critical 
exponent (Fig.\ \ref{22nu}). Actually, one should expect in those two
cases the behavior similar to one-channel model due to the fact that 
matrices ${\bf U}$ commute with each other. Our results support this 
expectation.

The next in the hierarchy of symmetries is the $SU(2)$ group 
(Eq.\ (\ref{umatr1}) 
with $\delta =0$) with three independent phases. (A choice of 
$2$ independent phases is not closed under successive 
transfer operations). 

The most general case of a network model corresponding to the Hamiltonian 
Eq.\ (\ref{pauli}) is that of SU(2) matrices on links and a complex 
transfer matrix (Eq.\ (\ref{first})) at nodes. 
The ${\bf T}$ matrix on links is then a $U(2)$ matrix, however only 
its $SU(2)$ phases are random while the phase $\delta$ in Eq. (\ref{umatr1}) 
is regular, having the values $\pm\alpha,\pm\beta$ periodically. 

Before presenting numerical results we discuss the symmetries for which
$\xi_{\infty}(\epsilon)$ is invariant. Note first that $\alpha
\rightarrow \alpha
+\pi$ or $\beta \rightarrow \beta +\pi$ is a symmetry, since one can 
shift even or odd columns by a constant phase. Shifting by $\pi$, 
results in $\alpha +\pi$
and $-\alpha+2\pi$ (the latter is equivalent to $-(\alpha+\pi)+\pi$) or to
$\beta +\pi$ and $-\beta +\pi$ (the latter is now equivalent to 
$-\beta-\pi$). Consider next the up down reflection symmetry which 
yields (after shifts by
$\pi$) the equivalence $(\alpha, \beta)\rightarrow (-\alpha, -\beta)$. 
The right/left reflection symmetry leads to $(\alpha, \beta)\rightarrow
(\beta +\pi/2, \alpha -\pi/2)$. Indeed, looking from right to left, 
the links with phases $(\beta, -\beta)$ come first (actually with
opposite signs since propagation is to the
left, but by the previous symmetry signs can be changed) 
while $(\alpha, -\alpha +\pi)$ come second; shifting by $\pm \pi$ 
this yields the stated equivalence. The
next symmetry is a property of the infinite 2D system:
if we rotate our system by $90^{\circ}$ and then shift 
the phases properly we get $(\alpha ,\beta )\rightarrow 
((\alpha +\beta +\pi )/2, (\alpha +\beta -\pi )/2)$, but due to the 
right/left symmetry this is just $((\alpha +\beta )/2, (\alpha +\beta
)/2)$. If this symmetry holds for our finite strips then it is 
sufficient to consider only the diagonals in the $(\alpha, \beta)$ 
plane. Note that all these symmetries 
hold for the spectra of the pure system, Eq. (14). Finally, we 
consider a symmetry which is due to the random $SU(2)$ phases. A shift 
of the link phase $\delta$ in Eq.\ (\ref{umatr1}) by $\pi$ is 
equivalent to a shift of $\pi$ in the $SU(2)$ phases $\delta_{1}$, 
$\delta_{2}$; since the latter are random the result is invariant. 
Thus in Fig.\ \ref{picture}a the phase $-\alpha+\pi$ can be replaced 
by $-\alpha$ and a column shift by $\pi /2$ yields $(\alpha, 
\beta )\rightarrow (\alpha +\pi /2, \beta )$ or similarly
$(\alpha, \beta )\rightarrow (\alpha , \beta +\pi /2)$. Thus with 
rotation symmetry it is sufficient to consider phases $(\alpha , 
\alpha )$ in the range $0<\alpha <\pi /4$.

We checked all these symmetries numerically. It turned out that all 
symmetries hold,
however, the rotational symmetry requires larger M and higher number of 
iterations than the $U(2)$ case, in particular for large 
$\alpha$, $\beta$. This
fact has probably the following explanation: the periodic  
phases $\alpha ,\beta $ introduce a new irrelevant length scale in 
the system. In order to go beyond this scale to find the symmetries 
one needs to investigate much longer systems (at least $300,000$ 
iterations in comparison with $60,000$ in the $U(2)$ case). The  
rotational symmetry was found for the largest system size $M=128$ 
for some particular phases. It is expected to hold for $M\rightarrow
\infty$.

The results for $(\alpha =-0.175, \beta =0.075)$ and 
$(\alpha =-0.005, \beta =-0.005)$
are shown in Figs.\ 
\ref{shaic175},\ref{shaic005all} respectively. The data exhibit single
peaks near $\epsilon =0$. The peculiar property of the data is that 
in the range between $\epsilon\approx -0.1$ and $\epsilon =0$ the 
renormalized localization length $\xi_{16}/8<\xi_{32}/16<\xi_{64}/32$
which indicates a band of extended states, however  
the data for $M=128$ shows $\xi_{64}/32>\xi_{128}/64$ which determines
these to be localized 
states. This is just another manifestation of the 
irrelevant length scale mentioned above.

We consider finally the case $\alpha =\beta =0$ in Figs.\ \ref{su2raw},
\ref{nu11}. Although
the pure case is singular, the disordered system is well behaved, the
data converges faster and there is no irrelevant length scale, i.e. 
the localized behavior of $\epsilon\neq 0$ states sets in already at 
$M=32$. We therefore consider $\alpha =\beta =0$ as the generic case
for $SU(2)$.

There is clearly a single extended state at $\epsilon =0$, 
or more precisely at least all states with $|\epsilon |>0.03$ are localized. 
Recall, that in the $U(2)$ case with $\Delta =0$ the extended states
are at $\epsilon \approx\pm 0.2$ well separated from $\epsilon =0$ 
(see Refs. {\onlinecite{rlee,rleeko,rwang}). 
Hence if there are extended states 
in the $SU(2)$ case near
$\epsilon =0$, they are extremely close to $\epsilon =0$ and 
therefore are not related to those of the $U(2)$ case as has been
proposed \cite{dkkl}.

The critical exponent in Fig.\ \ref{nu11} is $\nu\approx 1.1$. 
Thus the $SU(2)$
case is a new universality class. This is consistent with the 
occurrence of singular density of states and new value of 
$\sigma_{xx}$ found by Hikami et al. \cite{rhikami}

Finally, we consider an interpolation between $U(2)$ and $SU(2)$ by 
allowing the link phase ($\delta$ in Eq. 12) to be random in a 
restricted range 
of $p[-\pi ,\pi]$. On long scales the randomness may accumulate, 
allowing an $SU(2)$ behavior
only at $p=0$. However, it is known that a metal-insulator transition 
(in the absence of a magnetic field) occurs at a finite ratio of
spin-orbit to scalar randomness
\cite{simplectic}. Thus, it may be possible that $SU(2)$ behavior sets
in at a finite $p$. Fig.\ \ref{pdat} shows our data with $p=0.3$; the best fit 
for scaling form yields extended states at $\epsilon=\pm 0.16$ with 
$\nu=2.5$.  Since the extended states of the $U(2)$ case ($p=1$) are 
at $\epsilon \approx\pm 0.2$ \cite{rlee,rleeko,rwang} we conclude that
the two extended states approach each other as $p$ is reduced until 
below $p\approx 0.2$ they
merge and only $\epsilon=0$ correspond to an extended state (or states). The
range between $p\approx 0.2$ and $p\approx 0.05$ is difficult to
analyze since
the localized nature of the states near $\epsilon=0$ sets in only at $M=128$;
apparently there is an irrelevant length scale, similar to the one we had
above. Below $p\approx 0.05$ the behavior is close to that of the SU(2)
(i.e.$p=0$) case, i.e. Figs.\ \ref{su2raw},\ref{nu11}. Thus we have a $U(2)$ 
to $SU(2)$ phase 
transition at a finite $p$, $p\approx 0.2$.

We summarize our results in Table I. The table shows the different 
symmetries involved with their critical properties.   

\section{Discussion}
\label{concl}

We have studied two types of systems: (i) Non-degenerate states with random
mixing and random scalar potential, and (ii) degenerate states with randomness
only in the mixing terms. The system with non-degenerate states is relevant to
two types of experimentally studied cases. The first case is where 
non-degeneracy
is represented by Landau level splitting. We find that the lower Landau level
has a minimum as function of magnetic field, consistent with the floating
scenario \cite{rkhmel,rlaugh}. This result accounts for a transition
from a Hall liquid to an insulator at both high and low fields, as observed
experimentally \cite{rexp}.

The second case is where non-degeneracy is represented by the Zeeman spin
splitting. In this case the bare splitting $\Delta$ (which is 
$g^{*}\mu_B B/E_1$ in
the simplest case) can cross the value $\approx 0.5$ where level repulsion
crosses into level attraction. We predict therefore that the spin splitting of
the extended states is larger (smaller) than the bare Zeeman splitting at low
(high) fields. To estimate the field corresponding to $\Delta\approx
0.5$ we use
data \cite{Wei} on temperature scaling of the conductance peak in
$GaAs/Al_xGa_{1-x}As$ which shows that the low temperature scaling of the
$N=0\downarrow$ peak near $B=6 T$ is modified at a crossover temperature of
$\approx 0.3 K$. Analysis of tunneling through saddle point barriers
\cite{rour} shows that the crossover temperature is $T_1=\hbar U/(2\pi
m\omega_c)$ where $U (=U_x=U_y)$ is the barrier curvature. Since $T_1\ll
\omega_c \approx 120 K$ at $B=6 T$ (using an effective mass \cite{Lassing} of
$m/m_0\approx 0.07$) Eq. (4) yields $E_1\approx \hbar U/m\omega_c =2\pi
T_1\approx 2 K$ and $(\hbar U/m)^{1/2} \approx 15 K$.  Using the bulk value
\cite{dobers} $g^{*}=0.44$ we obtain $\Delta\approx 1$, i.e. level attraction is
predicted above $B\approx 4 T$ (where $\Delta\approx 0.5$).

Spin splitting of extended states can be directly probed by ESR; the extended
electromagnetic wave couples dominantly the extended states. Experimental data
\cite{dobers} on Landau levels $N=0, 1, 2$ shows a spin splitting which is
nonlinear in B and was fitted by a quadratic polynomial. Band structure
calculations \cite{Lassing,Lommer} show that at large fields the pure system
has a spin splitting of the form $\Delta E_0=g^{*}\mu _B B+\nu_N$, assuming
\cite{Lassing} $\nu_N\ll \omega_c$; (in the more recent calculation
\cite{Lommer} the linear form is valid for $N=0, B>6 T$). 

The nonlinearity is best seen in the $N=0$ data which spans $B=8-14$T and is
clearly nonlinear in the range $8-12$T. The band structure calculations
\cite{Lassing,Lommer} predict linear behavior, at least at $B>6$T. We propose
then that the deviation from linearity is due to localization effects, i.e.
level attraction. Note also that if indeed the ESR transition probes only
extended states, it accounts for the remarkably low linewidth \cite{dobers} of
$50$mT.

To fit the data we choose $\Delta E_0=g^* \mu_B B+\nu_0$ with 
$g^*m/2m_0=0.018$
and $\nu_0=0$ (i.e. for \cite{Lassing} $m/m_0=0.07$ we have $g^*=0.51$). The
parameter $\Delta$ of Fig. 3 is now $\Delta=\Delta E_0/E_1$, allowing the data
of Fig. 3 to be replotted as the splitting 
$\Delta E=E_1(\epsilon_2-\epsilon_1)$
in Fig. 4. The experimental data could also be fitted with a smaller $g^*$ by
increasing $\nu _0$, e.g. $g^*=0.44$ with $\nu _0=11$GHz. The splitting in
frequency units is given by $(U/m)^{1/2}=310$GHz while the field scale
is set by $\omega_c(U/m)^{1/2}=1.3B$(T). 

Fig. 4 shows nonlinear behavior with a decreasing slope for $B>8$T, 
in agreement
with the experimental data \cite{dobers}. Below $8$T we predict a different
trend with the energy splitting crossing the bare one at $\approx 4$T (Fig. 4)
and becoming larger than the bare splitting at $B<4$T. (Recall that the bare
splitting itself, which should be determined by band structure calculation, may
be nonlinear in $B$ for these lower fields \cite{Lassing,Lommer}). We propose
then that extending the ESR data to lower and higher fields can serve as a
unique tool for testing our predictions for level attraction and repulsion.

We have also studied the network model with random mixing of two degenerate
states, while the scalar potential is periodic (non-random). The random
part of the transfer matrix Eq. (12) is then an SU(2) matrix. We
believe that this model
corresponds to that of Hikami et al. \cite{rhikami} Eq. (13). There
are, however,
some differences: a) The continuum translation symmetry of the free particle
term in Eq. (13) is replaced by a periodic potential with extended
states in the
band of Eq. (14). This band resembles the continuum one for small $\alpha$,
$\beta$. b) The random fields $h_x$, $h_y$ of Eq. (12) correspond in 
our case to
a general $SU(2)$ matrix, i.e. we have also a random $h_z$ component. 
While Eq. (13) including a random $h_z$ was not explicitly 
studied, we  believe that the
noncommutative Pauli matrices $\sigma_x$, $\sigma_y$ generate $\sigma_z$ 
terms in the time evolution (or in perturbation theory) and the models
are therefore equivalent.

Our results for the SU(2) model show that for small but finite 
$\alpha$, $\beta$
a single extended state appears at $\epsilon=0$, though the 
localized nature of
nearby states appears only at unusually wide strips ($M=128$). This 
indicates that
the phases $\alpha$, $\beta$ produce a new (irrelevant) length scale which
renders numerical analysis more difficult. Remarkably, for the choice
$\alpha=\beta=0$ the irrelevant length scale disappears and we find a clear
single extended state at $\epsilon=0$. Finite size scaling is well 
obeyed (Fig.\ \ref{nu11}) and an exponent $\nu=1.1$ is found. Thus the $SU(2)$ 
network model forms a new universality class.

Our $SU(2)$ data are inconsistent with Lee's argument \cite{dkkl} and 
with the numerical study of Minakuchi and Hikami \cite{rjap}. Lee's argument is
qualitative and applies to slowly varying random fields 
${\bf h}({\bf r})$ while
in our case ${\bf h}({\bf r})$ is uncorrelated between unit cells. 
We do not
expect that this will change critical properties since correlations will vanish
along long loops in both cases. The numerical data of Minakuchi and Hikami, 
who study Eq. (13),
shows $3$ extended states. At present we cannot account
for this discrepancy with regards to the location of extended states; although
Eq. (13) is defined differently from our $SU(2)$ model, the two should be
equivalent, as discussed above.

Experimental realizations of the $SU(2)$ model correspond to systems 
with dominant
spin-orbit coupling. We predict that in a usual random system which has spin
split levels, increasing the strength of spin-orbit coupling would shift the
extended states so that they approach each other, until at some critical value
of the spin orbit coupling, the energies of these two extended states would
merge. We simulated this situation by allowing randomness in link 
phases to vary
in the range $p[-\pi ,\pi]$ range. We find that below 
$p\approx 0.2$ the system
behaves like an $SU(2)$ one, i.e. an extended state (or
states) at the single energy of the band center.

\begin{center}
{\bf ACKNOWLEDGMENTS} 
\end{center}
We wish to thank  
Prof. J. T. Chalker for valuable discussions, for sending his data to us
 and for hospitality during 
visit of one of us (V. K.). We also thank M. Feingold and Y. Meir for 
useful discussions. One of us (V. K.) also thanks Dr. B. Huckestein and 
Dr. D. Polyakov for valuable discussions.

\appendix
\section{Test for Sublocalization}
\label{subloc}

The simplest argument for deriving the $7/3$ exponent \cite{rmyl} 
is based on the 
semiclassical picture. An electron with energy $E\neq 0$ follows 
semiclassical constant potential contours \cite{rtrugman} on 
clusters of size of the percolation length $\xi_p$ which according to percolation theory
diverges as $\xi_p\sim |E|^{-4/3}$. An electron can propagate
to a distance $r\gg\xi_p$ by tunneling events between clusters, each
event reducing the wavefunction by a factor $\exp (-a|E|)$ and $a$ 
depends on potential barrier parameters. These 
tunneling events occur at saddle points in the potential which are on the 
infinite percolating cluster at $E=0$. Assuming that these saddle points 
are homogeneously distributed, i.e. their number is $r/\xi_p$, we find 
that the total wavefunction decay is 
\begin{equation}
\left[\exp (-a|E|)\right]^{r/\xi_p}=\exp (-r/\xi ).
\label{deriv73}
\end{equation}
Hence the localization length $\xi\sim\xi_p/|E|\sim |E|^{-7/3}$.

We wish to test the homogeneity assumption, i.e. whether the number of 
tunneling events is $r/\xi_p$. In the theory of percolation 
the infinite cluster at the critical energy 
has a backbone consisting of 
alternating sequence of singly connected ("red") bonds and 
multiply connected ("blue") bonds \cite{rred,rrred,rred1}. It is 
conceivable that tunneling occurs 
at "red" bonds of the infinite cluster, i.e. bonds whose elimination 
disconnects the infinite cluster. Analytic \cite{rred1} 
and numerical \cite{rred2} studies show that the number of "red" bonds 
$L_{red}$ between two points separated by a correlation length $\xi_p$ 
diverges as $E^{-1}$; therefore at the scales 
 $r=\xi_p\sim E^{-4/3}$  
we obtain $L_{BB}\sim r^{3/4}$. If we assume that the number of 
tunneling events is $(r/\xi_p)^{3/4}$, the derivation of the localization 
length Eq.\ (\ref{deriv73}) is now modified to  $\exp 
[(-aE)(r/\xi_p)^{3/4}]\sim\exp [-(r/\xi)^{3/4}]$. The consequences are 
that now the localization length exponent is modified 
$\xi\sim\ E^{-8/3}$, and more significantly that sublocalization 
appears with a weaker $\exp [-(r/\xi )^{3/4}]$ decay. 
To check this possibility consider how it 
affects finite size scaling and in particular the region of 
localized states \cite{rlamb}.
 Suppose that wave 
function on an $M\times M$ square decreases 
as $\exp [-(M/\tilde{\xi}_M)^{\beta}]$. Then at distance $r>M$, where $M$
 is the width of the strip, it 
decays as in a $1$D strip, i.e. as 
$\exp [-(M/\tilde {\xi}_M)^{\beta}(r/M)]=\exp[-r/\xi_M]$, and 
$\xi_M$ is 
the value that we calculate numerically; hence, $\xi_M=(\tilde 
{\xi}_M/M)^{\beta}M$. On 
on the other hand, by scaling hypothesis 
$\tilde {\xi}_M/M=f(\xi/M)$, which for 
$\xi\ll\ M$ yields $\tilde {\xi}_M/M=\xi/M$ and 
therefore 
$\xi_M/M=(\xi/M)^{\beta}$. Therefore an asymptotic tangent to the
 fitting curve of  
$\log (\xi_M/M)$ versus $\log (M/\xi )$ should have a slope 
$-\beta$ ($-3/4$ if the idea of "red" bonds is valid). We 
perform optimization procedure using our results  
as well as data from CC \cite{rchalkpr}. The fitting curve
(Fig.\ \ref{beta}) 
and its tangent in the range $\xi\ll M$ 
show conclusively 
that $\beta$ can not be distinguished from unity and 
therefore the localized 
wave functions decay exponentially.

\section{Mixing of Landau levels}

We wish to estimate the mixing probability of Landau levels on the links, i.e. 
far from saddle points. We consider then 
the tunneling rate between Landau levels in a 
weakly perturbed harmonic potential. 
We choose the potential on the links as 
$V(x,y)=(1/2)\tilde{U}(x^2+y^2)-(\tilde{U}/\lambda ) y^3$; $\lambda$ is a 
a measure of the correlation length of the potential. 
 We perform a
standard change of variables
\begin{equation}
{\bf\hat{\pi}}={\bf\hat{p}}-\frac{e{\bf\hat{A}}}{c},\ \       
\hat{X}=x-\frac{c\hat{\pi}_y}{eB},\    
\hat{Y}=y+\frac{c\hat{\pi}_x}{eB}.    
\label{cv}
\end{equation}
$\hat{X}$ and $\hat{Y}$ are slow guiding center coordinates and 
$\hat{\pi}_{x,y}$  are 
operators of fast cyclotron motion.
A Schr\"{o}dinger equation for the fast variables 
(choosing $c\hat{\pi}_x/eB=z$ as a coordinate and treating 
$\hat{X}$ and $\hat{Y}$ adiabatically) has a form  
\begin{eqnarray}
\left[\left(\frac{1}{2m}+\frac{\tilde{U}}{2m^2\omega_{c}^{2}}\right)
\left(\frac{d}{dz}+\frac{\tilde{U}m^2\omega_{c}X}{m\omega_{c}^{2}+
\tilde{U}}\right)^2
+\frac{m}{2}\left(\omega_{c}^{2}+\frac{\tilde{U}}
{m}-\frac{3\tilde{U}Y}{\lambda m}\right)
\left(z-\frac{\tilde{U}Y/m-(3\tilde{U}/\lambda m)Y^2}
{\omega_{c}^{2}+\tilde{U}/m-(6\tilde{U}/\lambda m)Y}
\right)^2+\frac{\tilde{U}}{\lambda} 
z^3\right]\nonumber\\
\times\psi_n(z;X,Y)=E_n(X,Y)\psi_n(z;X,Y).              
\label{srd} 
\end{eqnarray}
Eigenvalues of Eq. (\ref{srd}) can be interpreted \cite{rlevit} as local 
nonequidistant Landau levels $E_n(X,Y)$ depending on the classical guiding 
center coordinates.

We estimate the tunneling rate between neighbor Landau levels using 
Dykhne's formula \cite{rdyk,rmood}:
\begin{equation}
P_{n,n+1}\sim\exp [-(2/\hbar )\mbox{\rm Im}\int_{0}^{t_c}(E_{n+1}-E_n)dt],
\label{tun}
\end{equation}
where $E_{n+1}>E_n$ for all real times, and $t_c$ is a point in the 
complex time plane where they cross. We solve Eq. (\ref{srd}) keeping $X$ 
and $Y$ as fixed parameters and obtain to the first order in $1/\lambda$ 
\begin{equation}
E_{n+1}-E_n\approx\hbar\omega_c(1+\frac{\tilde{U}}
{m\omega_{c}^{2}}-\frac{3\tilde{U} 
Y/\lambda}{m\omega_{c}^{2}+\tilde{U}}).
\label{en}
\end{equation}
Note that we are considering quasi bound states near the local minimum 
of $V(x,y)$, i.e. $Y\ll \lambda$. We solve equations of motion for
$Y$ in a harmonic potential, substitute this solution into Eq. (\ref{en}) 
and find 
\begin{equation}
t_c=\frac{\pi}{2}\frac{m\omega_c}{\tilde{U}}\pm i\frac{m\omega_c}
{\tilde{U}}
\ln\frac{2\lambda (m\omega_{c}^{2}+\tilde{U})^2}{3m\tilde{U}\omega_{c}^{2}A},
\end{equation}
where $A$ is the amplitude of oscillations in the $Y$-direction. Note that 
$\tilde{U}A$ is a measure of the center coordinate energy which 
corresponds to the width of a Landau band; since this width is less 
than or on the order of $\hbar\omega_c$ we have $A\lesssim\sqrt{\hbar
\omega_{c}/\tilde{U}}=\ell\sqrt{m\omega_{c}^{2}/\tilde{U}}$ where 
$\ell$ is the cyclotron length. 
We find crossings at conjugate points as one 
expects for real Hamiltonians \cite{rmood}. Evaluating then 
Eq. (\ref{tun}) we 
find that the main contribution to the tunneling (in this adiabatic 
approximation) is independent on $n$ and proportional to
\begin{equation}
P_{n,n+1}\sim\exp\left(-\frac{m\omega_{c}^{2}+\tilde{U}}
{\tilde{U}}\ln\frac{2\lambda (m\omega_{c}^{2}
+\tilde{U})^2}
{3\ell\sqrt{m\tilde{U}}\omega_{c}^{3}}\right).
\label{last}
\end{equation}
For transitions from $n$ to $n+\Delta n$ the term in the exponent 
is multiplied by $\Delta n$.

\newpage
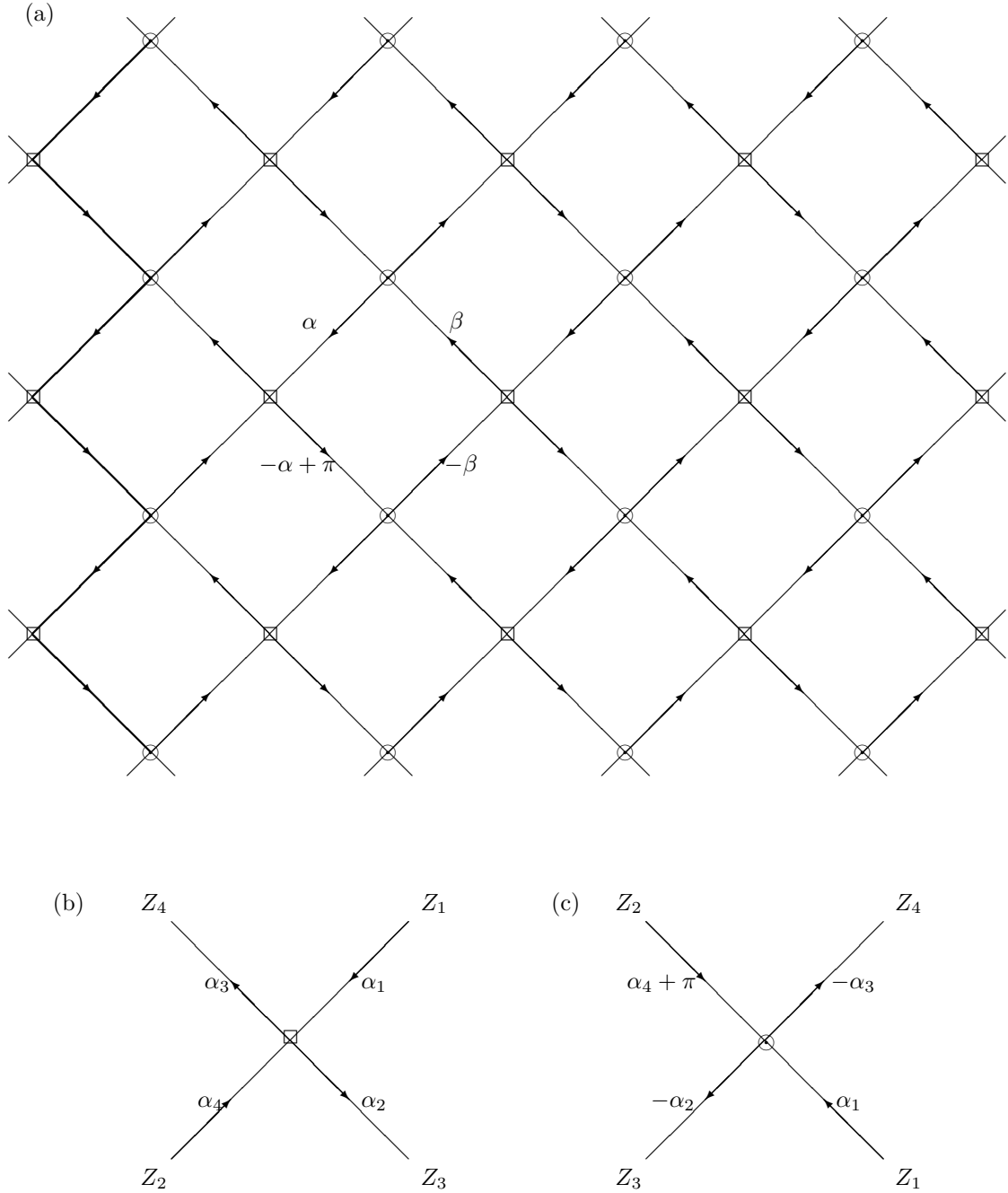
\begin{figure}
%
\begin{picture}(100,200)(-100,0)
{\thicklines
\multiput (-58.6,141.4)(0,-100){3}{\line(1,1){50}}
\multiput (-58.6,141.4)(0,-100){3}{\line(1,-1){50}}
}
\put (55,70){$\alpha$}
\put (37,10){$-\alpha +\pi$}
\put (117,70){$\beta$}
\put (115,10){$-\beta$}

\put(-62,200){(a)}
\multiput (-58.6,141.4)(100,0){4}{\line(1,1){50}}
\multiput (-62,138)(100,0){5}{$\Box$}
\multiput (-62,38)(100,0){5}{$\Box$}   
\multiput (-62,-62)(100,0){5}{$\Box$}
\multiput (-12.7,188.95)(100,0){4}{$\odot$}      
\multiput (-12.7,88.95)(100,0){4}{$\odot$} 
\multiput (-12.7,-11.05)(100,0){4}{$\odot$} 
\multiput (-12.7,-111.05)(100,0){4}{$\odot$} 
\multiput (-58.6,141.4)(0,-100){3}{\line(-1,1){10}} 
\multiput (341.4,141.4)(0,-100){3}{\line(1,1){10}} 
\multiput (-58.6,141.4)(0,-100){3}{\line(-1,-1){10}} 
\multiput (341.4,141.4)(0,-100){3}{\line(1,-1){10}} 
\multiput (-8.6,191.4)(100,0){4}{\vector(-1,-1){25}}
\multiput (-8.6,191.4)(100,0){4}{\line(1,1){10}}
\multiput (-8.6,191.4)(100,0){4}{\line(-1,1){10}}
\multiput (-58.6,141.4)(100,0){4}{\line(1,-1){50}}
\multiput (-58.6,141.4)(100,0){4}{\vector(1,-1){25}}
\multiput (41.4,141.4)(100,0){4}{\line(-1,1){50}}
\multiput (41.4,141.4)(100,0){4}{\vector(-1,1){25}}
\multiput (41.4,141.4)(100,0){4}{\line(-1,-1){50}}
\multiput (-8.6,91.6)(100,0){4}{\vector(1,1){25}}
\multiput (-58.6,41.4)(100,0){4}{\line(1,1){50}}
\multiput (-8.6,91.4)(100,0){4}{\vector(-1,-1){25}}
\multiput (-58.6,41.4)(100,0){4}{\line(1,-1){50}}
\multiput (-58.6,41.4)(100,0){4}{\vector(1,-1){25}}
\multiput (41.4,41.4)(100,0){4}{\line(-1,1){50}}
\multiput (41.4,41.4)(100,0){4}{\vector(-1,1){25}}
\multiput (41.4,41.4)(100,0){4}{\line(-1,-1){50}}
\multiput (-8.6,-8.4)(100,0){4}{\vector(1,1){25}}
\multiput (-58.6,-58.6)(100,0){4}{\line(1,1){50}}
\multiput (-8.6,-8.6)(100,0){4}{\vector(-1,-1){25}}
\multiput (-58.6,-58.6)(100,0){4}{\line(1,-1){50}}
\multiput (-58.6,-58.6)(100,0){4}{\vector(1,-1){25}}
\multiput (41.4,-58.6)(100,0){4}{\line(-1,1){50}}
\multiput (41.4,-58.6)(100,0){4}{\vector(-1,1){25}}
\multiput (41.4,-58.6)(100,0){4}{\line(-1,-1){50}}
\multiput (-8.6,-108.4)(100,0){4}{\vector(1,1){25}}
\multiput (-8.6,-108.4)(100,0){4}{\line(-1,-1){10}}
\multiput (-8.6,-108.4)(100,0){4}{\line(1,-1){10}}
\put (46.5,-232){$\Box$}
\put (50,-230){\line(1,1){50}}
\put (0,-280){\vector(1,1){25}}
\put (50,-230){\line(-1,1){50}}
\put (50,-230){\vector(1,-1){25}}
\put (50,-230){\line(1,-1){50}}
\put (50,-230){\vector(-1,1){25}}
\put (50,-230){\line(-1,-1){50}}
\put (100,-180){\vector(-1,-1){25}}
\put(-50,-175){(b)}
\put(160,-175){(c)}
\put (105,-175){$Z_1$}
\put (80,-208){$\alpha_1$}
\put (-13,-175){$Z_4$}
\put (14,-208){$\alpha_3$}
\put (-13,-291){$Z_2$}
\put (11,-258){$\alpha_4$}
\put (105,-291){$Z_3$}
\put (80,-258){$\alpha_2$}

\put (246.8,-233.5){$\odot$}
\put (250,-230){\line(1,1){50}}
\put (250,-230){\vector(1,1){25}}
\put (250,-230){\line(-1,1){50}}
\put (200,-180){\vector(1,-1){25}}
\put (250,-230){\line(1,-1){50}}
\put (300,-280){\vector(-1,1){25}}
\put (250,-230){\line(-1,-1){50}}
\put (250,-230){\vector(-1,-1){25}}
\put (305,-175){$Z_4$}
\put (278,-208){$-\alpha_3$}
\put (187,-175){$Z_2$}
\put (192,-208){$\alpha_4 +\pi$}
\put (187,-291){$Z_3$}
\put (202,-258){$-\alpha_2$}
\put (305,-291){$Z_1$}
\put (280,-258){$\alpha_1$}

\end{picture}
\vspace{11cm}
\caption{The network model. Arrows indicate direction of 
current flow. The nodes marked by $\circ$ and $\Box$ are related 
by a $90^{\circ}$ rotation. (a) The strip has $M_l$ links with
periodic boundary 
conditions ($M_l=6$ in the figure). In the two channel case each link 
represents two channels, i.e. $M=2M_l$ is the number of transverse channels. 
The thick line is a reference for measuring area in Section 4. (b) and (c)
Phase relations for two neighbor nodes.}
\label{picture}
\end{figure}

\newpage
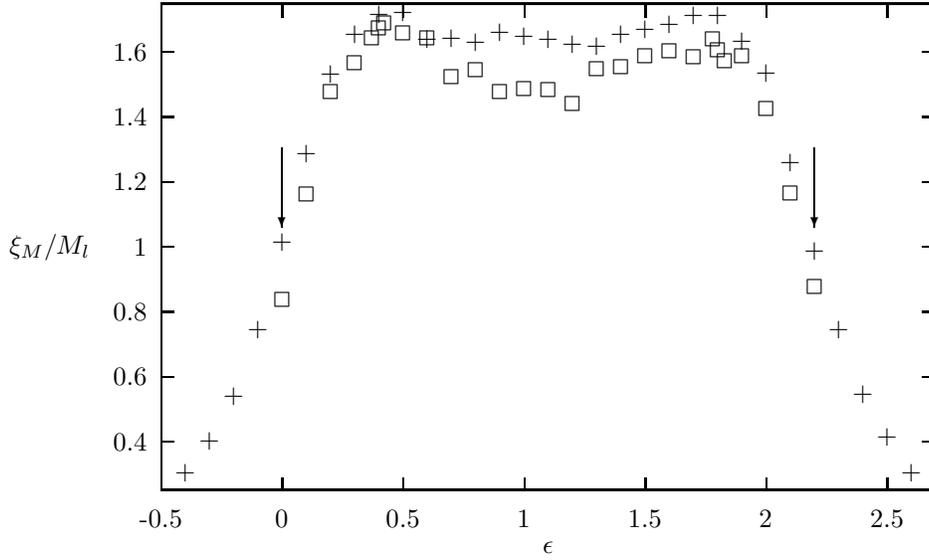
\begin{figure}
%
\setlength{\unitlength}{0.240900pt}
\ifx\plotpoint\undefined\newsavebox{\plotpoint}\fi
\sbox{\plotpoint}{\rule[-0.200pt]{0.400pt}{0.400pt}}%
\begin{picture}(1500,900)(0,0)
\font\gnuplot=cmr10 at 10pt
\gnuplot
\sbox{\plotpoint}{\rule[-0.200pt]{0.400pt}{0.400pt}}%
\put(220.0,189.0){\rule[-0.200pt]{4.818pt}{0.400pt}}
\put(198,189){\makebox(0,0)[r]{0.4}}
\put(1416.0,189.0){\rule[-0.200pt]{4.818pt}{0.400pt}}
\put(220.0,291.0){\rule[-0.200pt]{4.818pt}{0.400pt}}
\put(198,291){\makebox(0,0)[r]{0.6}}
\put(1416.0,291.0){\rule[-0.200pt]{4.818pt}{0.400pt}}
\put(220.0,393.0){\rule[-0.200pt]{4.818pt}{0.400pt}}
\put(198,393){\makebox(0,0)[r]{0.8}}
\put(1416.0,393.0){\rule[-0.200pt]{4.818pt}{0.400pt}}
\put(220.0,495.0){\rule[-0.200pt]{4.818pt}{0.400pt}}
\put(198,495){\makebox(0,0)[r]{1}}
\put(1416.0,495.0){\rule[-0.200pt]{4.818pt}{0.400pt}}
\put(220.0,597.0){\rule[-0.200pt]{4.818pt}{0.400pt}}
\put(198,597){\makebox(0,0)[r]{1.2}}
\put(1416.0,597.0){\rule[-0.200pt]{4.818pt}{0.400pt}}
\put(220.0,699.0){\rule[-0.200pt]{4.818pt}{0.400pt}}
\put(198,699){\makebox(0,0)[r]{1.4}}
\put(1416.0,699.0){\rule[-0.200pt]{4.818pt}{0.400pt}}
\put(220.0,801.0){\rule[-0.200pt]{4.818pt}{0.400pt}}
\put(198,801){\makebox(0,0)[r]{1.6}}
\put(1416.0,801.0){\rule[-0.200pt]{4.818pt}{0.400pt}}
\put(220.0,113.0){\rule[-0.200pt]{0.400pt}{4.818pt}}
\put(220,68){\makebox(0,0){-0.5}}
\put(220.0,857.0){\rule[-0.200pt]{0.400pt}{4.818pt}}
\put(410.0,113.0){\rule[-0.200pt]{0.400pt}{4.818pt}}
\put(410,68){\makebox(0,0){0}}
\put(410.0,857.0){\rule[-0.200pt]{0.400pt}{4.818pt}}
\put(600.0,113.0){\rule[-0.200pt]{0.400pt}{4.818pt}}
\put(600,68){\makebox(0,0){0.5}}
\put(600.0,857.0){\rule[-0.200pt]{0.400pt}{4.818pt}}
\put(790.0,113.0){\rule[-0.200pt]{0.400pt}{4.818pt}}
\put(790,68){\makebox(0,0){1}}
\put(790.0,857.0){\rule[-0.200pt]{0.400pt}{4.818pt}}
\put(980.0,113.0){\rule[-0.200pt]{0.400pt}{4.818pt}}
\put(980,68){\makebox(0,0){1.5}}
\put(980.0,857.0){\rule[-0.200pt]{0.400pt}{4.818pt}}
\put(1170.0,113.0){\rule[-0.200pt]{0.400pt}{4.818pt}}
\put(1170,68){\makebox(0,0){2}}
\put(1170.0,857.0){\rule[-0.200pt]{0.400pt}{4.818pt}}
\put(1360.0,113.0){\rule[-0.200pt]{0.400pt}{4.818pt}}
\put(1360,68){\makebox(0,0){2.5}}
\put(1360.0,857.0){\rule[-0.200pt]{0.400pt}{4.818pt}}
\put(220.0,113.0){\rule[-0.200pt]{292.934pt}{0.400pt}}
\put(1436.0,113.0){\rule[-0.200pt]{0.400pt}{184.048pt}}
\put(220.0,877.0){\rule[-0.200pt]{292.934pt}{0.400pt}}
\put(45,495){\makebox(0,0){$\xi_M/M_l$}}
\put(828,23){\makebox(0,0){$\epsilon$}}
\put(220.0,113.0){\rule[-0.200pt]{0.400pt}{184.048pt}}
\put(1398,140){\makebox(0,0){$+$}}
\put(1360,196){\makebox(0,0){$+$}}
\put(1322,264){\makebox(0,0){$+$}}
\put(1284,365){\makebox(0,0){$+$}}
\put(1246,489){\makebox(0,0){$+$}}
\put(1208,628){\makebox(0,0){$+$}}
\put(1170,768){\makebox(0,0){$+$}}
\put(1132,818){\makebox(0,0){$+$}}
\put(1094,859){\makebox(0,0){$+$}}
\put(1056,859){\makebox(0,0){$+$}}
\put(1018,844){\makebox(0,0){$+$}}
\put(980,837){\makebox(0,0){$+$}}
\put(942,829){\makebox(0,0){$+$}}
\put(904,811){\makebox(0,0){$+$}}
\put(866,813){\makebox(0,0){$+$}}
\put(828,822){\makebox(0,0){$+$}}
\put(790,826){\makebox(0,0){$+$}}
\put(752,832){\makebox(0,0){$+$}}
\put(714,817){\makebox(0,0){$+$}}
\put(676,823){\makebox(0,0){$+$}}
\put(638,822){\makebox(0,0){$+$}}
\put(600,863){\makebox(0,0){$+$}}
\put(562,861){\makebox(0,0){$+$}}
\put(524,829){\makebox(0,0){$+$}}
\put(486,766){\makebox(0,0){$+$}}
\put(448,642){\makebox(0,0){$+$}}
\put(410,502){\makebox(0,0){$+$}}
\put(372,365){\makebox(0,0){$+$}}
\put(334,261){\makebox(0,0){$+$}}
\put(296,190){\makebox(0,0){$+$}}
\put(258,140){\makebox(0,0){$+$}}

\put(1246,550){\vector(0,-1){25}}
\put(1246,650){\line(0,-1){100}} 
\put(410,550){\vector(0,-1){25}}
\put(410,650){\line(0,-1){100}}

\put(1246,431){\raisebox{-.8pt}{\makebox(0,0){$\Box$}}}
\put(1208,578){\raisebox{-.8pt}{\makebox(0,0){$\Box$}}}
\put(1170,710){\raisebox{-.8pt}{\makebox(0,0){$\Box$}}}
\put(1132,794){\raisebox{-.8pt}{\makebox(0,0){$\Box$}}}
\put(1105,785){\raisebox{-.8pt}{\makebox(0,0){$\Box$}}}
\put(1094,803){\raisebox{-.8pt}{\makebox(0,0){$\Box$}}}
\put(1086,820){\raisebox{-.8pt}{\makebox(0,0){$\Box$}}}
\put(1056,792){\raisebox{-.8pt}{\makebox(0,0){$\Box$}}}
\put(1018,802){\raisebox{-.8pt}{\makebox(0,0){$\Box$}}}
\put(980,794){\raisebox{-.8pt}{\makebox(0,0){$\Box$}}}
\put(942,776){\raisebox{-.8pt}{\makebox(0,0){$\Box$}}}
\put(904,773){\raisebox{-.8pt}{\makebox(0,0){$\Box$}}}
\put(866,718){\raisebox{-.8pt}{\makebox(0,0){$\Box$}}}
\put(828,741){\raisebox{-.8pt}{\makebox(0,0){$\Box$}}}
\put(790,742){\raisebox{-.8pt}{\makebox(0,0){$\Box$}}}
\put(752,737){\raisebox{-.8pt}{\makebox(0,0){$\Box$}}}
\put(714,772){\raisebox{-.8pt}{\makebox(0,0){$\Box$}}}
\put(676,761){\raisebox{-.8pt}{\makebox(0,0){$\Box$}}}
\put(638,821){\raisebox{-.8pt}{\makebox(0,0){$\Box$}}}
\put(600,829){\raisebox{-.8pt}{\makebox(0,0){$\Box$}}}
\put(570,845){\raisebox{-.8pt}{\makebox(0,0){$\Box$}}}
\put(562,838){\raisebox{-.8pt}{\makebox(0,0){$\Box$}}}
\put(551,821){\raisebox{-.8pt}{\makebox(0,0){$\Box$}}}
\put(524,782){\raisebox{-.8pt}{\makebox(0,0){$\Box$}}}
\put(486,737){\raisebox{-.8pt}{\makebox(0,0){$\Box$}}}
\put(448,577){\raisebox{-.8pt}{\makebox(0,0){$\Box$}}}
\put(410,411){\raisebox{-.8pt}{\makebox(0,0){$\Box$}}}
\end{picture}

\caption{ Renormalized localization length $\xi_M /M$
as a function of $\epsilon$  
for $\Delta =2.2$. $+$ symbols correspond  
to $M=32$ and $\Box$ to $M=64$ system widths. Arrows point to the 
location of extended states in the absence of level mixing. The 
energies of extended states (near the peaks of $\xi_{M}/M_{l}$) are 
closer than the arrow positions, demonstrating level attraction.}
\label{e32}
\end{figure}

\newpage
\vspace*{-6cm}
\begin{figure}
\centerline{ \psfig{figure=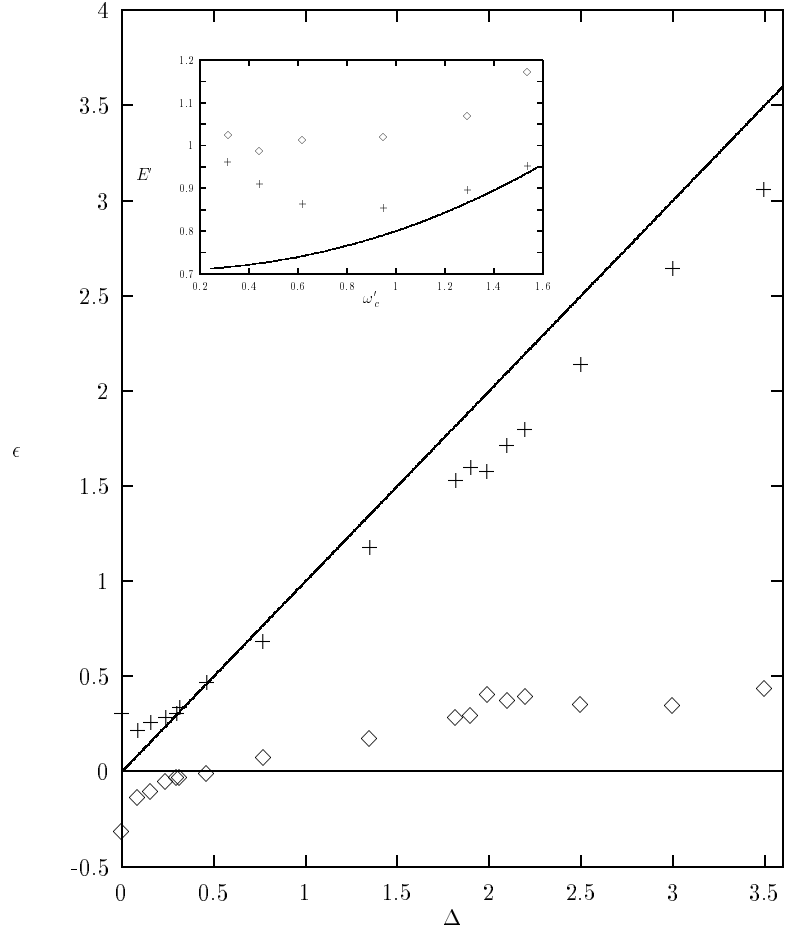,width=8in,angle=0}}
\caption{Critical values $\epsilon$ as functions of $\Delta$. Full lines 
are the bare extended states $\epsilon =0$ and $\epsilon =\Delta$ are
shown. Inset:  Energy of the lowest extended state as a function of 
magnetic field  with full level mixing ($\Diamond$) and with 
reduced mixing ($+$). Full line is the lowest bare extended
state. Here $E'=E(2m/U)^{1/2}$, $\omega '_c=\omega_c(2m/U)^{1/2}$.}
\label{eps}
\end{figure}

\begin{figure}
\input{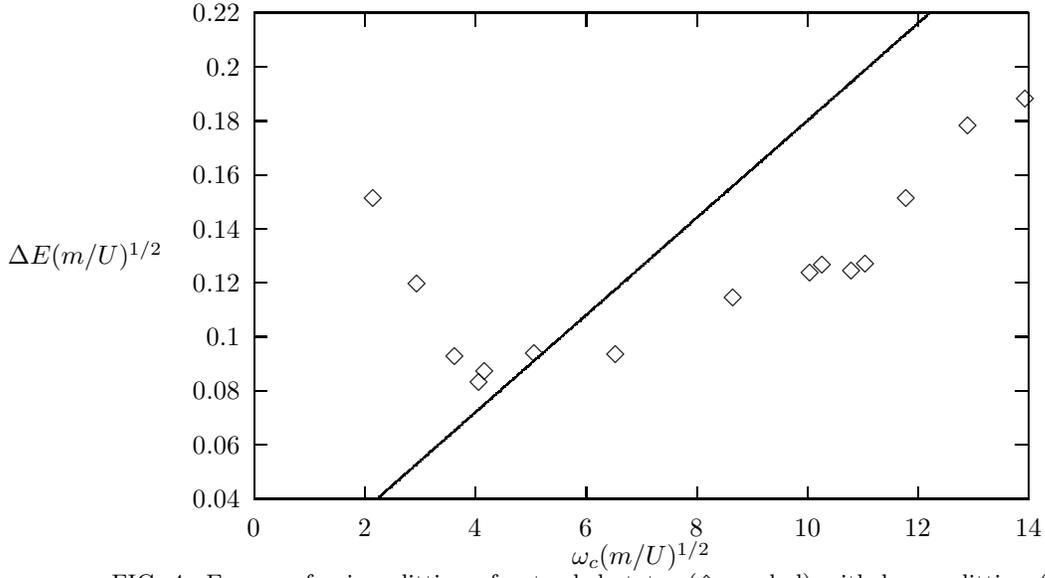}
\caption{Energy of spin splitting of extended states ($\Diamond$ symbol) 
with bare splitting (straight line) of $g^* \mu _B B=0.018\omega_c$. 
$GaAs/Al_xGa_{1-x}As$
data {\protect\cite{dobers}} fits our results with
$(U/m)^{1/2}=310$GHz 
and the field
scale is $\omega_c (U/m)^{1/2}\approx 1.3 B$(T).}
\label{diff}
\end{figure}

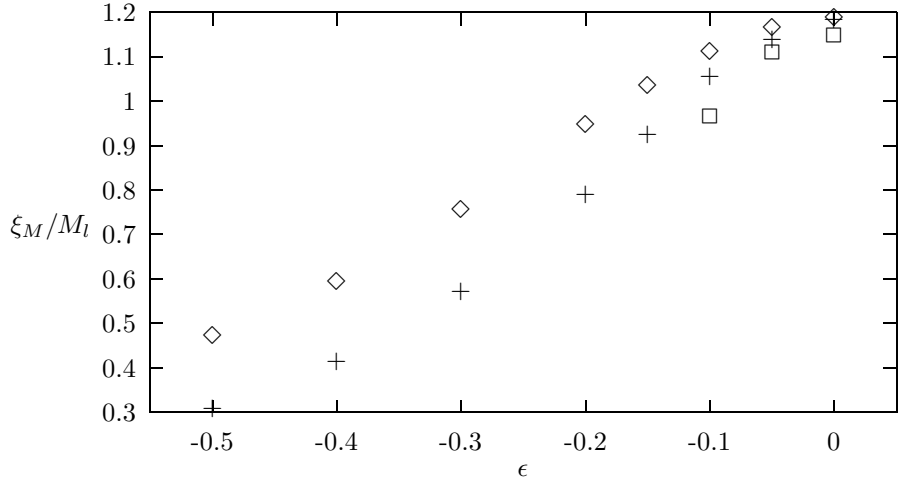
\begin{figure}
%
\setlength{\unitlength}{0.240900pt}
\ifx\plotpoint\undefined\newsavebox{\plotpoint}\fi
\sbox{\plotpoint}{\rule[-0.175pt]{0.350pt}{0.350pt}}%
\begin{picture}(1500,900)(0,0)
\sbox{\plotpoint}{\rule[-0.175pt]{0.350pt}{0.350pt}}%
\put(264,158){\rule[-0.175pt]{4.818pt}{0.350pt}}
\put(242,158){\makebox(0,0)[r]{0.3}}
\put(1416,158){\rule[-0.175pt]{4.818pt}{0.350pt}}
\put(264,228){\rule[-0.175pt]{4.818pt}{0.350pt}}
\put(242,228){\makebox(0,0)[r]{0.4}}
\put(1416,228){\rule[-0.175pt]{4.818pt}{0.350pt}}
\put(264,298){\rule[-0.175pt]{4.818pt}{0.350pt}}
\put(242,298){\makebox(0,0)[r]{0.5}}
\put(1416,298){\rule[-0.175pt]{4.818pt}{0.350pt}}
\put(264,368){\rule[-0.175pt]{4.818pt}{0.350pt}}
\put(242,368){\makebox(0,0)[r]{0.6}}
\put(1416,368){\rule[-0.175pt]{4.818pt}{0.350pt}}
\put(264,438){\rule[-0.175pt]{4.818pt}{0.350pt}}
\put(242,438){\makebox(0,0)[r]{0.7}}
\put(1416,438){\rule[-0.175pt]{4.818pt}{0.350pt}}
\put(264,507){\rule[-0.175pt]{4.818pt}{0.350pt}}
\put(242,507){\makebox(0,0)[r]{0.8}}
\put(1416,507){\rule[-0.175pt]{4.818pt}{0.350pt}}
\put(264,577){\rule[-0.175pt]{4.818pt}{0.350pt}}
\put(242,577){\makebox(0,0)[r]{0.9}}
\put(1416,577){\rule[-0.175pt]{4.818pt}{0.350pt}}
\put(264,647){\rule[-0.175pt]{4.818pt}{0.350pt}}
\put(242,647){\makebox(0,0)[r]{1}}
\put(1416,647){\rule[-0.175pt]{4.818pt}{0.350pt}}
\put(264,717){\rule[-0.175pt]{4.818pt}{0.350pt}}
\put(242,717){\makebox(0,0)[r]{1.1}}
\put(1416,717){\rule[-0.175pt]{4.818pt}{0.350pt}}
\put(264,787){\rule[-0.175pt]{4.818pt}{0.350pt}}
\put(242,787){\makebox(0,0)[r]{1.2}}
\put(1416,787){\rule[-0.175pt]{4.818pt}{0.350pt}}
\put(362,158){\rule[-0.175pt]{0.350pt}{4.818pt}}
\put(362,113){\makebox(0,0){-0.5}}
\put(362,767){\rule[-0.175pt]{0.350pt}{4.818pt}}
\put(557,158){\rule[-0.175pt]{0.350pt}{4.818pt}}
\put(557,113){\makebox(0,0){-0.4}}
\put(557,767){\rule[-0.175pt]{0.350pt}{4.818pt}}
\put(752,158){\rule[-0.175pt]{0.350pt}{4.818pt}}
\put(752,113){\makebox(0,0){-0.3}}
\put(752,767){\rule[-0.175pt]{0.350pt}{4.818pt}}
\put(948,158){\rule[-0.175pt]{0.350pt}{4.818pt}}
\put(948,113){\makebox(0,0){-0.2}}
\put(948,767){\rule[-0.175pt]{0.350pt}{4.818pt}}
\put(1143,158){\rule[-0.175pt]{0.350pt}{4.818pt}}
\put(1143,113){\makebox(0,0){-0.1}}
\put(1143,767){\rule[-0.175pt]{0.350pt}{4.818pt}}
\put(1338,158){\rule[-0.175pt]{0.350pt}{4.818pt}}
\put(1338,113){\makebox(0,0){0}}
\put(1338,767){\rule[-0.175pt]{0.350pt}{4.818pt}}
\put(264,158){\rule[-0.175pt]{282.335pt}{0.350pt}}
\put(1436,158){\rule[-0.175pt]{0.350pt}{151.526pt}}
\put(264,787){\rule[-0.175pt]{282.335pt}{0.350pt}}
\put(45,450){\makebox(0,0)[l]{\shortstack{$\xi_M/M_l$}}}
\put(850,68){\makebox(0,0){$\epsilon$}}
\put(264,158){\rule[-0.175pt]{0.350pt}{151.526pt}}
\put(362,279){\raisebox{-1.2pt}{\makebox(0,0){$\Diamond$}}}
\put(557,364){\raisebox{-1.2pt}{\makebox(0,0){$\Diamond$}}}
\put(752,478){\raisebox{-1.2pt}{\makebox(0,0){$\Diamond$}}}
\put(948,611){\raisebox{-1.2pt}{\makebox(0,0){$\Diamond$}}}
\put(1045,673){\raisebox{-1.2pt}{\makebox(0,0){$\Diamond$}}}
\put(1143,726){\raisebox{-1.2pt}{\makebox(0,0){$\Diamond$}}}
\put(1241,764){\raisebox{-1.2pt}{\makebox(0,0){$\Diamond$}}}
\put(1338,780){\raisebox{-1.2pt}{\makebox(0,0){$\Diamond$}}}
\put(362,164){\makebox(0,0){$+$}}
\put(557,238){\makebox(0,0){$+$}}
\put(752,349){\makebox(0,0){$+$}}
\put(948,501){\makebox(0,0){$+$}}
\put(1045,596){\makebox(0,0){$+$}}
\put(1143,687){\makebox(0,0){$+$}}
\put(1241,745){\makebox(0,0){$+$}}
\put(1338,776){\makebox(0,0){$+$}}
\put(1143,624){\raisebox{-1.2pt}{\makebox(0,0){$\Box$}}}
\put(1241,725){\raisebox{-1.2pt}{\makebox(0,0){$\Box$}}}
\put(1338,751){\raisebox{-1.2pt}{\makebox(0,0){$\Box$}}}
\end{picture}
\caption{Renormalized localization length $\xi_M/M_l$ as function of 
energy $\epsilon$ for a system with $SO(2)$ mixing 
between the levels. $\Diamond$ symbols correspond  
to $M=16$, $+$ to $M=32$ and $\Box$ to $M=64$ system widths.}   
\label{u1raw}
\end{figure}

\begin{figure}
%
\setlength{\unitlength}{0.240900pt}
\ifx\plotpoint\undefined\newsavebox{\plotpoint}\fi
\sbox{\plotpoint}{\rule[-0.175pt]{0.350pt}{0.350pt}}%
\begin{picture}(1500,900)(0,0)
\sbox{\plotpoint}{\rule[-0.175pt]{0.350pt}{0.350pt}}%
\put(264,158){\rule[-0.175pt]{4.818pt}{0.350pt}}
\put(242,158){\makebox(0,0)[r]{0.2}}
\put(1416,158){\rule[-0.175pt]{4.818pt}{0.350pt}}
\put(264,272){\rule[-0.175pt]{4.818pt}{0.350pt}}
\put(242,272){\makebox(0,0)[r]{0.4}}
\put(1416,272){\rule[-0.175pt]{4.818pt}{0.350pt}}
\put(264,387){\rule[-0.175pt]{4.818pt}{0.350pt}}
\put(242,387){\makebox(0,0)[r]{0.6}}
\put(1416,387){\rule[-0.175pt]{4.818pt}{0.350pt}}
\put(264,501){\rule[-0.175pt]{4.818pt}{0.350pt}}
\put(242,501){\makebox(0,0)[r]{0.8}}
\put(1416,501){\rule[-0.175pt]{4.818pt}{0.350pt}}
\put(264,615){\rule[-0.175pt]{4.818pt}{0.350pt}}
\put(242,615){\makebox(0,0)[r]{1}}
\put(1416,615){\rule[-0.175pt]{4.818pt}{0.350pt}}
\put(264,730){\rule[-0.175pt]{4.818pt}{0.350pt}}
\put(242,730){\makebox(0,0)[r]{1.2}}
\put(1416,730){\rule[-0.175pt]{4.818pt}{0.350pt}}
\put(293,158){\rule[-0.175pt]{0.350pt}{4.818pt}}
\put(293,113){\makebox(0,0){0}}
\put(293,767){\rule[-0.175pt]{0.350pt}{4.818pt}}
\put(578,158){\rule[-0.175pt]{0.350pt}{4.818pt}}
\put(578,113){\makebox(0,0){0.5}}
\put(578,767){\rule[-0.175pt]{0.350pt}{4.818pt}}
\put(864,158){\rule[-0.175pt]{0.350pt}{4.818pt}}
\put(864,113){\makebox(0,0){1}}
\put(864,767){\rule[-0.175pt]{0.350pt}{4.818pt}}
\put(1150,158){\rule[-0.175pt]{0.350pt}{4.818pt}}
\put(1150,113){\makebox(0,0){1.5}}
\put(1150,767){\rule[-0.175pt]{0.350pt}{4.818pt}}
\put(1436,158){\rule[-0.175pt]{0.350pt}{4.818pt}}
\put(1436,113){\makebox(0,0){2}}
\put(1436,767){\rule[-0.175pt]{0.350pt}{4.818pt}}
\put(264,158){\rule[-0.175pt]{282.335pt}{0.350pt}}
\put(1436,158){\rule[-0.175pt]{0.350pt}{151.526pt}}
\put(264,787){\rule[-0.175pt]{282.335pt}{0.350pt}}
\put(45,450){\makebox(0,0)[l]{\shortstack{$\xi_M/M_l$}}}
\put(850,68){\makebox(0,0){$-\epsilon M^{1/\nu}$}}
\put(264,158){\rule[-0.175pt]{0.350pt}{151.526pt}}
\put(1028,314){\raisebox{-1.2pt}{\makebox(0,0){$\Diamond$}}}
\put(881,384){\raisebox{-1.2pt}{\makebox(0,0){$\Diamond$}}}
\put(734,477){\raisebox{-1.2pt}{\makebox(0,0){$\Diamond$}}}
\put(587,586){\raisebox{-1.2pt}{\makebox(0,0){$\Diamond$}}}
\put(513,636){\raisebox{-1.2pt}{\makebox(0,0){$\Diamond$}}}
\put(440,680){\raisebox{-1.2pt}{\makebox(0,0){$\Diamond$}}}
\put(366,711){\raisebox{-1.2pt}{\makebox(0,0){$\Diamond$}}}
\put(293,724){\raisebox{-1.2pt}{\makebox(0,0){$\Diamond$}}}
\put(1301,220){\makebox(0,0){$+$}}
\put(1099,280){\makebox(0,0){$+$}}
\put(897,372){\makebox(0,0){$+$}}
\put(696,496){\makebox(0,0){$+$}}
\put(595,573){\makebox(0,0){$+$}}
\put(494,648){\makebox(0,0){$+$}}
\put(393,696){\makebox(0,0){$+$}}
\put(293,721){\makebox(0,0){$+$}}
\put(569,596){\raisebox{-1.2pt}{\makebox(0,0){$\Box$}}}
\put(431,679){\raisebox{-1.2pt}{\makebox(0,0){$\Box$}}}
\put(293,700){\raisebox{-1.2pt}{\makebox(0,0){$\Box$}}}
\put(1301,220){\makebox(0,0){$\times$}}
\put(1099,282){\makebox(0,0){$\times$}}
\put(897,377){\makebox(0,0){$\times$}}
\put(696,511){\makebox(0,0){$\times$}}
\put(494,659){\makebox(0,0){$\times$}}
\put(293,718){\makebox(0,0){$\times$}}
\put(1322,225){\makebox(0,0){$\triangle$}}
\put(1175,263){\makebox(0,0){$\triangle$}}
\put(1028,315){\makebox(0,0){$\triangle$}}
\put(881,385){\makebox(0,0){$\triangle$}}
\put(734,477){\makebox(0,0){$\triangle$}}
\put(587,583){\makebox(0,0){$\triangle$}}
\put(440,689){\makebox(0,0){$\triangle$}}
\put(293,731){\makebox(0,0){$\triangle$}}
\put(1121,279){\makebox(0,0){$\star$}}
\put(845,422){\makebox(0,0){$\star$}}
\put(569,610){\makebox(0,0){$\star$}}
\put(293,743){\makebox(0,0){$\star$}}
\end{picture}
\caption{Fit of raw data from Fig. {\protect\ref{u1raw}} for a system with 
$SO(2)$ and for a system with $U(1)\times SO(2)$ mixing between the levels: 
$\xi_M/M_l$ versus $\epsilon M_{l}^{1/\nu}$ with 
$\nu =2.2$. $\Diamond$ symbols correspond  
to $M=16$, $+$ to $M=32$ and $\Box$ to $M=64$ system widths for $SO(2)$ 
case; $\times$ symbols correspond  
to $M=16$, $\triangle$ to $M=32$ and $\star$ to $M=64$ system widths for 
$U(1)\times SO(2)$ case.}
\label{22nu}
\end{figure}
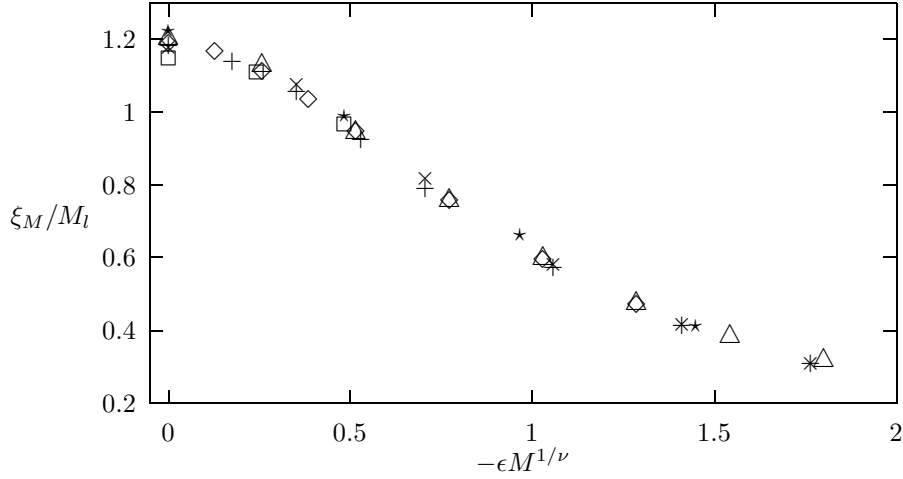

\begin{figure}
%
\setlength{\unitlength}{0.240900pt}
\ifx\plotpoint\undefined\newsavebox{\plotpoint}\fi
\sbox{\plotpoint}{\rule[-0.200pt]{0.400pt}{0.400pt}}%
\begin{picture}(1500,900)(0,0)
\font\gnuplot=cmr10 at 10pt
\gnuplot
\sbox{\plotpoint}{\rule[-0.200pt]{0.400pt}{0.400pt}}%
\put(220.0,113.0){\rule[-0.200pt]{292.934pt}{0.400pt}}
\put(220.0,113.0){\rule[-0.200pt]{4.818pt}{0.400pt}}
\put(198,113){\makebox(0,0)[r]{0}}
\put(1416.0,113.0){\rule[-0.200pt]{4.818pt}{0.400pt}}
\put(220.0,292.0){\rule[-0.200pt]{4.818pt}{0.400pt}}
\put(198,292){\makebox(0,0)[r]{0.5}}
\put(1416.0,292.0){\rule[-0.200pt]{4.818pt}{0.400pt}}
\put(220.0,472.0){\rule[-0.200pt]{4.818pt}{0.400pt}}
\put(198,472){\makebox(0,0)[r]{1}}
\put(1416.0,472.0){\rule[-0.200pt]{4.818pt}{0.400pt}}
\put(220.0,651.0){\rule[-0.200pt]{4.818pt}{0.400pt}}
\put(198,651){\makebox(0,0)[r]{1.5}}
\put(1416.0,651.0){\rule[-0.200pt]{4.818pt}{0.400pt}}
\put(220.0,830.0){\rule[-0.200pt]{4.818pt}{0.400pt}}
\put(198,830){\makebox(0,0)[r]{2}}
\put(1416.0,830.0){\rule[-0.200pt]{4.818pt}{0.400pt}}
\put(275.0,113.0){\rule[-0.200pt]{0.400pt}{4.818pt}}
\put(275,68){\makebox(0,0){-1}}
\put(275.0,857.0){\rule[-0.200pt]{0.400pt}{4.818pt}}
\put(496.0,113.0){\rule[-0.200pt]{0.400pt}{4.818pt}}
\put(496,68){\makebox(0,0){-0.8}}
\put(496.0,857.0){\rule[-0.200pt]{0.400pt}{4.818pt}}
\put(717.0,113.0){\rule[-0.200pt]{0.400pt}{4.818pt}}
\put(717,68){\makebox(0,0){-0.6}}
\put(717.0,857.0){\rule[-0.200pt]{0.400pt}{4.818pt}}
\put(939.0,113.0){\rule[-0.200pt]{0.400pt}{4.818pt}}
\put(939,68){\makebox(0,0){-0.4}}
\put(939.0,857.0){\rule[-0.200pt]{0.400pt}{4.818pt}}
\put(1160.0,113.0){\rule[-0.200pt]{0.400pt}{4.818pt}}
\put(1160,68){\makebox(0,0){-0.2}}
\put(1160.0,857.0){\rule[-0.200pt]{0.400pt}{4.818pt}}
\put(1381.0,113.0){\rule[-0.200pt]{0.400pt}{4.818pt}}
\put(1381,68){\makebox(0,0){0}}
\put(1381.0,857.0){\rule[-0.200pt]{0.400pt}{4.818pt}}
\put(220.0,113.0){\rule[-0.200pt]{292.934pt}{0.400pt}}
\put(1436.0,113.0){\rule[-0.200pt]{0.400pt}{184.048pt}}
\put(220.0,877.0){\rule[-0.200pt]{292.934pt}{0.400pt}}
\put(45,495){\makebox(0,0){$\xi_M/M_l$}}
\put(828,23){\makebox(0,0){$\epsilon$}}
\put(220.0,113.0){\rule[-0.200pt]{0.400pt}{184.048pt}}
\put(1381,830){\raisebox{-.8pt}{\makebox(0,0){$\Diamond$}}}
\put(1359,830){\raisebox{-.8pt}{\makebox(0,0){$\Diamond$}}}
\put(1325,808){\raisebox{-.8pt}{\makebox(0,0){$\Diamond$}}}
\put(1303,785){\raisebox{-.8pt}{\makebox(0,0){$\Diamond$}}}
\put(1270,738){\raisebox{-.8pt}{\makebox(0,0){$\Diamond$}}}
\put(1215,647){\raisebox{-.8pt}{\makebox(0,0){$\Diamond$}}}
\put(1160,550){\raisebox{-.8pt}{\makebox(0,0){$\Diamond$}}}
\put(1049,387){\raisebox{-.8pt}{\makebox(0,0){$\Diamond$}}}
\put(939,292){\raisebox{-.8pt}{\makebox(0,0){$\Diamond$}}}
\put(828,241){\raisebox{-.8pt}{\makebox(0,0){$\Diamond$}}}
\put(717,212){\raisebox{-.8pt}{\makebox(0,0){$\Diamond$}}}
\put(607,195){\raisebox{-.8pt}{\makebox(0,0){$\Diamond$}}}
\put(496,183){\raisebox{-.8pt}{\makebox(0,0){$\Diamond$}}}
\put(386,174){\raisebox{-.8pt}{\makebox(0,0){$\Diamond$}}}
\put(275,168){\raisebox{-.8pt}{\makebox(0,0){$\Diamond$}}}
\put(1381,862){\makebox(0,0){$+$}}
\put(1359,858){\makebox(0,0){$+$}}
\put(1325,830){\makebox(0,0){$+$}}
\put(1303,799){\makebox(0,0){$+$}}
\put(1270,737){\makebox(0,0){$+$}}
\put(1215,615){\makebox(0,0){$+$}}
\put(1160,497){\makebox(0,0){$+$}}
\put(1049,319){\makebox(0,0){$+$}}
\put(939,229){\makebox(0,0){$+$}}
\put(828,188){\makebox(0,0){$+$}}
\put(717,168){\makebox(0,0){$+$}}
\put(607,157){\makebox(0,0){$+$}}
\put(496,150){\makebox(0,0){$+$}}
\put(386,145){\makebox(0,0){$+$}}
\put(275,141){\makebox(0,0){$+$}}
\put(1381,862){\raisebox{-.8pt}{\makebox(0,0){$\Box$}}}
\put(1359,853){\raisebox{-.8pt}{\makebox(0,0){$\Box$}}}
\put(1325,822){\raisebox{-.8pt}{\makebox(0,0){$\Box$}}}
\put(1303,797){\raisebox{-.8pt}{\makebox(0,0){$\Box$}}}
\put(1270,720){\raisebox{-.8pt}{\makebox(0,0){$\Box$}}}
\put(1215,583){\raisebox{-.8pt}{\makebox(0,0){$\Box$}}}
\put(939,183){\raisebox{-.8pt}{\makebox(0,0){$\Box$}}}
\put(1381,851){\makebox(0,0){$\times$}}
\end{picture}

\caption{Renormalized localization length $\xi_M/M_l$ as function of 
energy $\epsilon$ for a system with $SU(2)$ matrices on links and 
and a complex transfer matrix at nodes with 
small ($\alpha =-0.175,\beta =0.075$) phases after 480,000
iterations. $\Diamond$ symbols correspond  
to $M=16$, $+$ to $M=32$, $\Box$ to $M=64$ and $\times$ to $M=128$ 
system widths.}
\label{shaic175}
\end{figure}
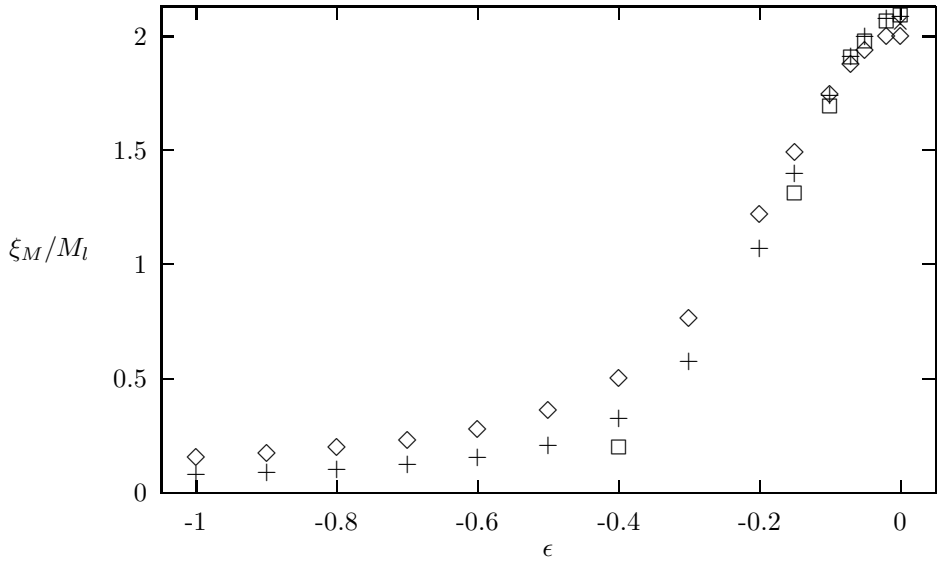

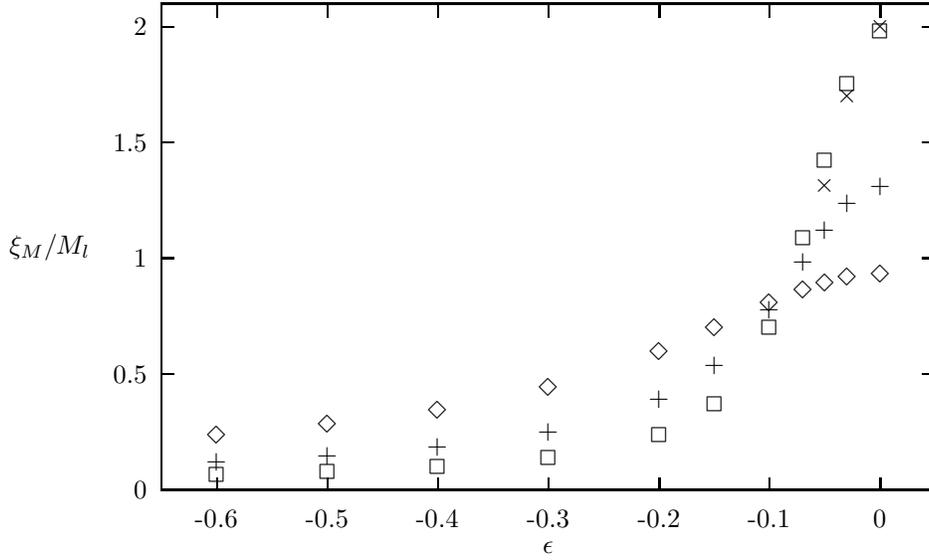
\begin{figure}
%
\setlength{\unitlength}{0.240900pt}
\ifx\plotpoint\undefined\newsavebox{\plotpoint}\fi
\sbox{\plotpoint}{\rule[-0.200pt]{0.400pt}{0.400pt}}%
\begin{picture}(1500,900)(0,0)
\font\gnuplot=cmr10 at 10pt
\gnuplot
\sbox{\plotpoint}{\rule[-0.200pt]{0.400pt}{0.400pt}}%
\put(220.0,113.0){\rule[-0.200pt]{292.934pt}{0.400pt}}
\put(220.0,113.0){\rule[-0.200pt]{4.818pt}{0.400pt}}
\put(198,113){\makebox(0,0)[r]{0}}
\put(1416.0,113.0){\rule[-0.200pt]{4.818pt}{0.400pt}}
\put(220.0,295.0){\rule[-0.200pt]{4.818pt}{0.400pt}}
\put(198,295){\makebox(0,0)[r]{0.5}}
\put(1416.0,295.0){\rule[-0.200pt]{4.818pt}{0.400pt}}
\put(220.0,477.0){\rule[-0.200pt]{4.818pt}{0.400pt}}
\put(198,477){\makebox(0,0)[r]{1}}
\put(1416.0,477.0){\rule[-0.200pt]{4.818pt}{0.400pt}}
\put(220.0,659.0){\rule[-0.200pt]{4.818pt}{0.400pt}}
\put(198,659){\makebox(0,0)[r]{1.5}}
\put(1416.0,659.0){\rule[-0.200pt]{4.818pt}{0.400pt}}
\put(220.0,841.0){\rule[-0.200pt]{4.818pt}{0.400pt}}
\put(198,841){\makebox(0,0)[r]{2}}
\put(1416.0,841.0){\rule[-0.200pt]{4.818pt}{0.400pt}}
\put(307.0,113.0){\rule[-0.200pt]{0.400pt}{4.818pt}}
\put(307,68){\makebox(0,0){-0.6}}
\put(307.0,857.0){\rule[-0.200pt]{0.400pt}{4.818pt}}
\put(481.0,113.0){\rule[-0.200pt]{0.400pt}{4.818pt}}
\put(481,68){\makebox(0,0){-0.5}}
\put(481.0,857.0){\rule[-0.200pt]{0.400pt}{4.818pt}}
\put(654.0,113.0){\rule[-0.200pt]{0.400pt}{4.818pt}}
\put(654,68){\makebox(0,0){-0.4}}
\put(654.0,857.0){\rule[-0.200pt]{0.400pt}{4.818pt}}
\put(828.0,113.0){\rule[-0.200pt]{0.400pt}{4.818pt}}
\put(828,68){\makebox(0,0){-0.3}}
\put(828.0,857.0){\rule[-0.200pt]{0.400pt}{4.818pt}}
\put(1002.0,113.0){\rule[-0.200pt]{0.400pt}{4.818pt}}
\put(1002,68){\makebox(0,0){-0.2}}
\put(1002.0,857.0){\rule[-0.200pt]{0.400pt}{4.818pt}}
\put(1175.0,113.0){\rule[-0.200pt]{0.400pt}{4.818pt}}
\put(1175,68){\makebox(0,0){-0.1}}
\put(1175.0,857.0){\rule[-0.200pt]{0.400pt}{4.818pt}}
\put(1349.0,113.0){\rule[-0.200pt]{0.400pt}{4.818pt}}
\put(1349,68){\makebox(0,0){0}}
\put(1349.0,857.0){\rule[-0.200pt]{0.400pt}{4.818pt}}
\put(220.0,113.0){\rule[-0.200pt]{292.934pt}{0.400pt}}
\put(1436.0,113.0){\rule[-0.200pt]{0.400pt}{184.048pt}}
\put(220.0,877.0){\rule[-0.200pt]{292.934pt}{0.400pt}}
\put(45,495){\makebox(0,0){$\xi_M/M_l$}}
\put(828,23){\makebox(0,0){$\epsilon$}}
\put(220.0,113.0){\rule[-0.200pt]{0.400pt}{184.048pt}}
\put(1349,451){\raisebox{-.8pt}{\makebox(0,0){$\Diamond$}}}
\put(1297,446){\raisebox{-.8pt}{\makebox(0,0){$\Diamond$}}}
\put(1262,438){\raisebox{-.8pt}{\makebox(0,0){$\Diamond$}}}
\put(1228,427){\raisebox{-.8pt}{\makebox(0,0){$\Diamond$}}}
\put(1175,406){\raisebox{-.8pt}{\makebox(0,0){$\Diamond$}}}
\put(1089,367){\raisebox{-.8pt}{\makebox(0,0){$\Diamond$}}}
\put(1002,330){\raisebox{-.8pt}{\makebox(0,0){$\Diamond$}}}
\put(828,274){\raisebox{-.8pt}{\makebox(0,0){$\Diamond$}}}
\put(654,238){\raisebox{-.8pt}{\makebox(0,0){$\Diamond$}}}
\put(481,215){\raisebox{-.8pt}{\makebox(0,0){$\Diamond$}}}
\put(307,198){\raisebox{-.8pt}{\makebox(0,0){$\Diamond$}}}
\put(1349,590){\makebox(0,0){$+$}}
\put(1297,564){\makebox(0,0){$+$}}
\put(1262,521){\makebox(0,0){$+$}}
\put(1228,471){\makebox(0,0){$+$}}
\put(1175,397){\makebox(0,0){$+$}}
\put(1089,308){\makebox(0,0){$+$}}
\put(1002,256){\makebox(0,0){$+$}}
\put(828,204){\makebox(0,0){$+$}}
\put(654,180){\makebox(0,0){$+$}}
\put(481,166){\makebox(0,0){$+$}}
\put(307,157){\makebox(0,0){$+$}}
\put(1349,832){\raisebox{-.8pt}{\makebox(0,0){$\Box$}}}
\put(1297,750){\raisebox{-.8pt}{\makebox(0,0){$\Box$}}}
\put(1262,630){\raisebox{-.8pt}{\makebox(0,0){$\Box$}}}
\put(1228,507){\raisebox{-.8pt}{\makebox(0,0){$\Box$}}}
\put(1175,367){\raisebox{-.8pt}{\makebox(0,0){$\Box$}}}
\put(1089,247){\raisebox{-.8pt}{\makebox(0,0){$\Box$}}}
\put(1002,198){\raisebox{-.8pt}{\makebox(0,0){$\Box$}}}
\put(828,162){\raisebox{-.8pt}{\makebox(0,0){$\Box$}}}
\put(654,148){\raisebox{-.8pt}{\makebox(0,0){$\Box$}}}
\put(481,140){\raisebox{-.8pt}{\makebox(0,0){$\Box$}}}
\put(307,136){\raisebox{-.8pt}{\makebox(0,0){$\Box$}}}
\put(1349,842){\makebox(0,0){$\times$}}
\put(1297,732){\makebox(0,0){$\times$}}
\put(1262,591){\makebox(0,0){$\times$}}
\end{picture}

\caption{Renormalized localization length $\xi_M/M_l$ as function of 
energy $\epsilon$ for a system with $SU(2)$ matrices on links and 
and a complex transfer matrix at nodes with 
small ($\alpha =\beta =-0.005$) phases after 480,000
iterations. $\Diamond$ symbols correspond  
to $M=16$, $+$ to $M=32$, $\Box$ to $M=64$ and $\times$ to $M=128$ 
system widths.}
\label{shaic005all}
\end{figure}

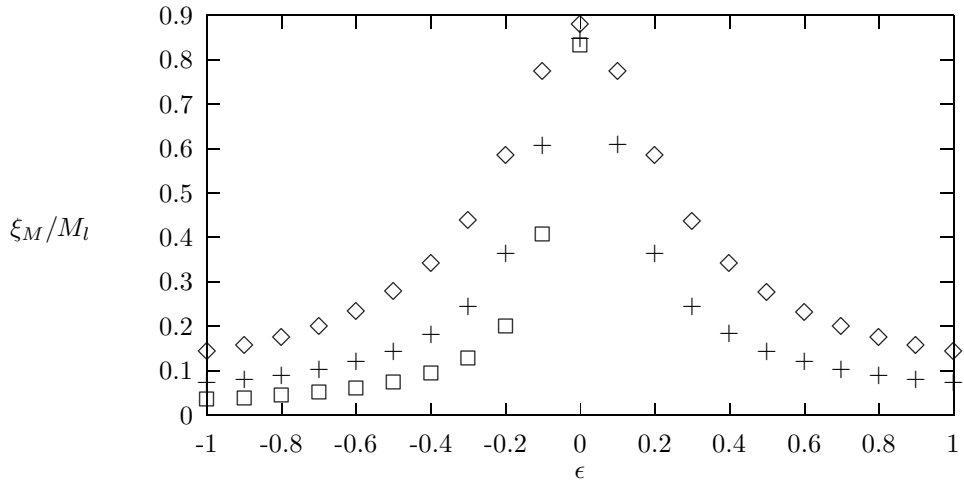
\begin{figure}
%
\setlength{\unitlength}{0.240900pt}
\ifx\plotpoint\undefined\newsavebox{\plotpoint}\fi
\sbox{\plotpoint}{\rule[-0.175pt]{0.350pt}{0.350pt}}%
\begin{picture}(1500,900)(0,0)
\sbox{\plotpoint}{\rule[-0.175pt]{0.350pt}{0.350pt}}%
\put(264,158){\rule[-0.175pt]{282.335pt}{0.350pt}}
\put(264,158){\rule[-0.175pt]{4.818pt}{0.350pt}}
\put(242,158){\makebox(0,0)[r]{0}}
\put(1416,158){\rule[-0.175pt]{4.818pt}{0.350pt}}
\put(264,228){\rule[-0.175pt]{4.818pt}{0.350pt}}
\put(242,228){\makebox(0,0)[r]{0.1}}
\put(1416,228){\rule[-0.175pt]{4.818pt}{0.350pt}}
\put(264,298){\rule[-0.175pt]{4.818pt}{0.350pt}}
\put(242,298){\makebox(0,0)[r]{0.2}}
\put(1416,298){\rule[-0.175pt]{4.818pt}{0.350pt}}
\put(264,368){\rule[-0.175pt]{4.818pt}{0.350pt}}
\put(242,368){\makebox(0,0)[r]{0.3}}
\put(1416,368){\rule[-0.175pt]{4.818pt}{0.350pt}}
\put(264,438){\rule[-0.175pt]{4.818pt}{0.350pt}}
\put(242,438){\makebox(0,0)[r]{0.4}}
\put(1416,438){\rule[-0.175pt]{4.818pt}{0.350pt}}
\put(264,507){\rule[-0.175pt]{4.818pt}{0.350pt}}
\put(242,507){\makebox(0,0)[r]{0.5}}
\put(1416,507){\rule[-0.175pt]{4.818pt}{0.350pt}}
\put(264,577){\rule[-0.175pt]{4.818pt}{0.350pt}}
\put(242,577){\makebox(0,0)[r]{0.6}}
\put(1416,577){\rule[-0.175pt]{4.818pt}{0.350pt}}
\put(264,647){\rule[-0.175pt]{4.818pt}{0.350pt}}
\put(242,647){\makebox(0,0)[r]{0.7}}
\put(1416,647){\rule[-0.175pt]{4.818pt}{0.350pt}}
\put(264,717){\rule[-0.175pt]{4.818pt}{0.350pt}}
\put(242,717){\makebox(0,0)[r]{0.8}}
\put(1416,717){\rule[-0.175pt]{4.818pt}{0.350pt}}
\put(264,787){\rule[-0.175pt]{4.818pt}{0.350pt}}
\put(242,787){\makebox(0,0)[r]{0.9}}
\put(1416,787){\rule[-0.175pt]{4.818pt}{0.350pt}}
\put(264,158){\rule[-0.175pt]{0.350pt}{4.818pt}}
\put(264,113){\makebox(0,0){-1}}
\put(264,767){\rule[-0.175pt]{0.350pt}{4.818pt}}
\put(381,158){\rule[-0.175pt]{0.350pt}{4.818pt}}
\put(381,113){\makebox(0,0){-0.8}}
\put(381,767){\rule[-0.175pt]{0.350pt}{4.818pt}}
\put(498,158){\rule[-0.175pt]{0.350pt}{4.818pt}}
\put(498,113){\makebox(0,0){-0.6}}
\put(498,767){\rule[-0.175pt]{0.350pt}{4.818pt}}
\put(616,158){\rule[-0.175pt]{0.350pt}{4.818pt}}
\put(616,113){\makebox(0,0){-0.4}}
\put(616,767){\rule[-0.175pt]{0.350pt}{4.818pt}}
\put(733,158){\rule[-0.175pt]{0.350pt}{4.818pt}}
\put(733,113){\makebox(0,0){-0.2}}
\put(733,767){\rule[-0.175pt]{0.350pt}{4.818pt}}
\put(850,158){\rule[-0.175pt]{0.350pt}{4.818pt}}
\put(850,113){\makebox(0,0){0}}
\put(850,767){\rule[-0.175pt]{0.350pt}{4.818pt}}
\put(967,158){\rule[-0.175pt]{0.350pt}{4.818pt}}
\put(967,113){\makebox(0,0){0.2}}
\put(967,767){\rule[-0.175pt]{0.350pt}{4.818pt}}
\put(1084,158){\rule[-0.175pt]{0.350pt}{4.818pt}}
\put(1084,113){\makebox(0,0){0.4}}
\put(1084,767){\rule[-0.175pt]{0.350pt}{4.818pt}}
\put(1202,158){\rule[-0.175pt]{0.350pt}{4.818pt}}
\put(1202,113){\makebox(0,0){0.6}}
\put(1202,767){\rule[-0.175pt]{0.350pt}{4.818pt}}
\put(1319,158){\rule[-0.175pt]{0.350pt}{4.818pt}}
\put(1319,113){\makebox(0,0){0.8}}
\put(1319,767){\rule[-0.175pt]{0.350pt}{4.818pt}}
\put(1436,158){\rule[-0.175pt]{0.350pt}{4.818pt}}
\put(1436,113){\makebox(0,0){1}}
\put(1436,767){\rule[-0.175pt]{0.350pt}{4.818pt}}
\put(264,158){\rule[-0.175pt]{282.335pt}{0.350pt}}
\put(1436,158){\rule[-0.175pt]{0.350pt}{151.526pt}}
\put(264,787){\rule[-0.175pt]{282.335pt}{0.350pt}}
\put(-45,450){\makebox(0,0)[l]{\shortstack{$\xi_M/M_l$}}}
\put(850,68){\makebox(0,0){$\epsilon$}}
\put(264,158){\rule[-0.175pt]{0.350pt}{151.526pt}}
\put(264,259){\raisebox{-1.2pt}{\makebox(0,0){$\Diamond$}}}
\put(323,269){\raisebox{-1.2pt}{\makebox(0,0){$\Diamond$}}}
\put(381,282){\raisebox{-1.2pt}{\makebox(0,0){$\Diamond$}}}
\put(440,299){\raisebox{-1.2pt}{\makebox(0,0){$\Diamond$}}}
\put(498,322){\raisebox{-1.2pt}{\makebox(0,0){$\Diamond$}}}
\put(557,353){\raisebox{-1.2pt}{\makebox(0,0){$\Diamond$}}}
\put(616,397){\raisebox{-1.2pt}{\makebox(0,0){$\Diamond$}}}
\put(674,465){\raisebox{-1.2pt}{\makebox(0,0){$\Diamond$}}}
\put(733,567){\raisebox{-1.2pt}{\makebox(0,0){$\Diamond$}}}
\put(791,698){\raisebox{-1.2pt}{\makebox(0,0){$\Diamond$}}}
\put(850,772){\raisebox{-1.2pt}{\makebox(0,0){$\Diamond$}}}
\put(909,698){\raisebox{-1.2pt}{\makebox(0,0){$\Diamond$}}}
\put(967,567){\raisebox{-1.2pt}{\makebox(0,0){$\Diamond$}}}
\put(1026,464){\raisebox{-1.2pt}{\makebox(0,0){$\Diamond$}}}
\put(1084,397){\raisebox{-1.2pt}{\makebox(0,0){$\Diamond$}}}
\put(1143,352){\raisebox{-1.2pt}{\makebox(0,0){$\Diamond$}}}
\put(1202,321){\raisebox{-1.2pt}{\makebox(0,0){$\Diamond$}}}
\put(1260,299){\raisebox{-1.2pt}{\makebox(0,0){$\Diamond$}}}
\put(1319,282){\raisebox{-1.2pt}{\makebox(0,0){$\Diamond$}}}
\put(1377,269){\raisebox{-1.2pt}{\makebox(0,0){$\Diamond$}}}
\put(1436,259){\raisebox{-1.2pt}{\makebox(0,0){$\Diamond$}}}
\put(264,209){\makebox(0,0){$+$}}
\put(323,214){\makebox(0,0){$+$}}
\put(381,221){\makebox(0,0){$+$}}
\put(440,230){\makebox(0,0){$+$}}
\put(498,242){\makebox(0,0){$+$}}
\put(557,259){\makebox(0,0){$+$}}
\put(616,285){\makebox(0,0){$+$}}
\put(674,329){\makebox(0,0){$+$}}
\put(733,412){\makebox(0,0){$+$}}
\put(791,582){\makebox(0,0){$+$}}
\put(850,750){\makebox(0,0){$+$}}
\put(909,584){\makebox(0,0){$+$}}
\put(967,413){\makebox(0,0){$+$}}
\put(1026,329){\makebox(0,0){$+$}}
\put(1084,286){\makebox(0,0){$+$}}
\put(1143,259){\makebox(0,0){$+$}}
\put(1202,242){\makebox(0,0){$+$}}
\put(1260,230){\makebox(0,0){$+$}}
\put(1319,221){\makebox(0,0){$+$}}
\put(1377,214){\makebox(0,0){$+$}}
\put(1436,209){\makebox(0,0){$+$}}
\put(264,184){\raisebox{-1.2pt}{\makebox(0,0){$\Box$}}}
\put(323,186){\raisebox{-1.2pt}{\makebox(0,0){$\Box$}}}
\put(381,190){\raisebox{-1.2pt}{\makebox(0,0){$\Box$}}}
\put(440,195){\raisebox{-1.2pt}{\makebox(0,0){$\Box$}}}
\put(498,201){\raisebox{-1.2pt}{\makebox(0,0){$\Box$}}}
\put(557,210){\raisebox{-1.2pt}{\makebox(0,0){$\Box$}}}
\put(616,224){\raisebox{-1.2pt}{\makebox(0,0){$\Box$}}}
\put(674,248){\raisebox{-1.2pt}{\makebox(0,0){$\Box$}}}
\put(733,299){\raisebox{-1.2pt}{\makebox(0,0){$\Box$}}}
\put(791,443){\raisebox{-1.2pt}{\makebox(0,0){$\Box$}}}
\put(850,739){\raisebox{-1.2pt}{\makebox(0,0){$\Box$}}}
\end{picture}
\caption{Renormalized localization length $\xi_M/M_l$ as function of 
energy $\epsilon$ for a system with $SU(2)$ mixing 
between the levels. $\Diamond$ symbols correspond  
to $M=16$, $+$ to $M=32$ and $\Box$ to $M=64$ system widths.}   
\label{su2raw}
\end{figure}

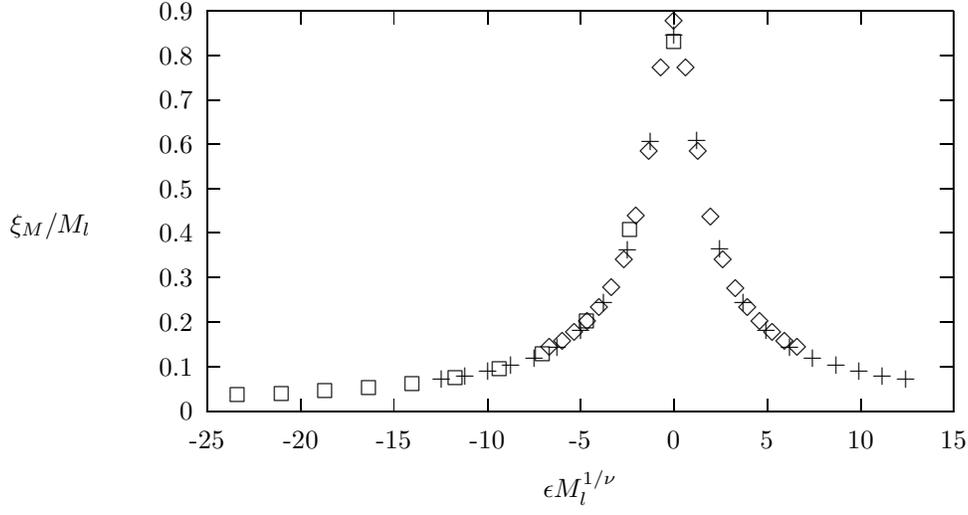
\begin{figure}
%
\setlength{\unitlength}{0.240900pt}
\ifx\plotpoint\undefined\newsavebox{\plotpoint}\fi
\sbox{\plotpoint}{\rule[-0.175pt]{0.350pt}{0.350pt}}%
\begin{picture}(1500,900)(0,0)
\sbox{\plotpoint}{\rule[-0.175pt]{0.350pt}{0.350pt}}%
\put(264,158){\rule[-0.175pt]{282.335pt}{0.350pt}}
\put(264,158){\rule[-0.175pt]{4.818pt}{0.350pt}}
\put(242,158){\makebox(0,0)[r]{0}}
\put(1416,158){\rule[-0.175pt]{4.818pt}{0.350pt}}
\put(264,228){\rule[-0.175pt]{4.818pt}{0.350pt}}
\put(242,228){\makebox(0,0)[r]{0.1}}
\put(1416,228){\rule[-0.175pt]{4.818pt}{0.350pt}}
\put(264,298){\rule[-0.175pt]{4.818pt}{0.350pt}}
\put(242,298){\makebox(0,0)[r]{0.2}}
\put(1416,298){\rule[-0.175pt]{4.818pt}{0.350pt}}
\put(264,368){\rule[-0.175pt]{4.818pt}{0.350pt}}
\put(242,368){\makebox(0,0)[r]{0.3}}
\put(1416,368){\rule[-0.175pt]{4.818pt}{0.350pt}}
\put(264,438){\rule[-0.175pt]{4.818pt}{0.350pt}}
\put(242,438){\makebox(0,0)[r]{0.4}}
\put(1416,438){\rule[-0.175pt]{4.818pt}{0.350pt}}
\put(264,507){\rule[-0.175pt]{4.818pt}{0.350pt}}
\put(242,507){\makebox(0,0)[r]{0.5}}
\put(1416,507){\rule[-0.175pt]{4.818pt}{0.350pt}}
\put(264,577){\rule[-0.175pt]{4.818pt}{0.350pt}}
\put(242,577){\makebox(0,0)[r]{0.6}}
\put(1416,577){\rule[-0.175pt]{4.818pt}{0.350pt}}
\put(264,647){\rule[-0.175pt]{4.818pt}{0.350pt}}
\put(242,647){\makebox(0,0)[r]{0.7}}
\put(1416,647){\rule[-0.175pt]{4.818pt}{0.350pt}}
\put(264,717){\rule[-0.175pt]{4.818pt}{0.350pt}}
\put(242,717){\makebox(0,0)[r]{0.8}}
\put(1416,717){\rule[-0.175pt]{4.818pt}{0.350pt}}
\put(264,787){\rule[-0.175pt]{4.818pt}{0.350pt}}
\put(242,787){\makebox(0,0)[r]{0.9}}
\put(1416,787){\rule[-0.175pt]{4.818pt}{0.350pt}}
\put(264,158){\rule[-0.175pt]{0.350pt}{4.818pt}}
\put(264,113){\makebox(0,0){-25}}
\put(264,767){\rule[-0.175pt]{0.350pt}{4.818pt}}
\put(411,158){\rule[-0.175pt]{0.350pt}{4.818pt}}
\put(411,113){\makebox(0,0){-20}}
\put(411,767){\rule[-0.175pt]{0.350pt}{4.818pt}}
\put(557,158){\rule[-0.175pt]{0.350pt}{4.818pt}}
\put(557,113){\makebox(0,0){-15}}
\put(557,767){\rule[-0.175pt]{0.350pt}{4.818pt}}
\put(704,158){\rule[-0.175pt]{0.350pt}{4.818pt}}
\put(704,113){\makebox(0,0){-10}}
\put(704,767){\rule[-0.175pt]{0.350pt}{4.818pt}}
\put(850,158){\rule[-0.175pt]{0.350pt}{4.818pt}}
\put(850,113){\makebox(0,0){-5}}
\put(850,767){\rule[-0.175pt]{0.350pt}{4.818pt}}
\put(997,158){\rule[-0.175pt]{0.350pt}{4.818pt}}
\put(997,113){\makebox(0,0){0}}
\put(997,767){\rule[-0.175pt]{0.350pt}{4.818pt}}
\put(1143,158){\rule[-0.175pt]{0.350pt}{4.818pt}}
\put(1143,113){\makebox(0,0){5}}
\put(1143,767){\rule[-0.175pt]{0.350pt}{4.818pt}}
\put(1290,158){\rule[-0.175pt]{0.350pt}{4.818pt}}
\put(1290,113){\makebox(0,0){10}}
\put(1290,767){\rule[-0.175pt]{0.350pt}{4.818pt}}
\put(1436,158){\rule[-0.175pt]{0.350pt}{4.818pt}}
\put(1436,113){\makebox(0,0){15}}
\put(1436,767){\rule[-0.175pt]{0.350pt}{4.818pt}}
\put(264,158){\rule[-0.175pt]{282.335pt}{0.350pt}}
\put(1436,158){\rule[-0.175pt]{0.350pt}{151.526pt}}
\put(264,787){\rule[-0.175pt]{282.335pt}{0.350pt}}
\put(-45,450){\makebox(0,0)[l]{\shortstack{$\xi_M/M_l$}}}
\put(850,38){\makebox(0,0){$\epsilon M_{l}^{1/\nu}$}}
\put(264,158){\rule[-0.175pt]{0.350pt}{151.526pt}}
\put(802,259){\raisebox{-1.2pt}{\makebox(0,0){$\Diamond$}}}
\put(822,269){\raisebox{-1.2pt}{\makebox(0,0){$\Diamond$}}}
\put(841,282){\raisebox{-1.2pt}{\makebox(0,0){$\Diamond$}}}
\put(861,299){\raisebox{-1.2pt}{\makebox(0,0){$\Diamond$}}}
\put(880,322){\raisebox{-1.2pt}{\makebox(0,0){$\Diamond$}}}
\put(899,353){\raisebox{-1.2pt}{\makebox(0,0){$\Diamond$}}}
\put(919,397){\raisebox{-1.2pt}{\makebox(0,0){$\Diamond$}}}
\put(938,465){\raisebox{-1.2pt}{\makebox(0,0){$\Diamond$}}}
\put(958,567){\raisebox{-1.2pt}{\makebox(0,0){$\Diamond$}}}
\put(977,698){\raisebox{-1.2pt}{\makebox(0,0){$\Diamond$}}}
\put(997,772){\raisebox{-1.2pt}{\makebox(0,0){$\Diamond$}}}
\put(1016,698){\raisebox{-1.2pt}{\makebox(0,0){$\Diamond$}}}
\put(1035,567){\raisebox{-1.2pt}{\makebox(0,0){$\Diamond$}}}
\put(1055,464){\raisebox{-1.2pt}{\makebox(0,0){$\Diamond$}}}
\put(1074,397){\raisebox{-1.2pt}{\makebox(0,0){$\Diamond$}}}
\put(1094,352){\raisebox{-1.2pt}{\makebox(0,0){$\Diamond$}}}
\put(1113,321){\raisebox{-1.2pt}{\makebox(0,0){$\Diamond$}}}
\put(1132,299){\raisebox{-1.2pt}{\makebox(0,0){$\Diamond$}}}
\put(1152,282){\raisebox{-1.2pt}{\makebox(0,0){$\Diamond$}}}
\put(1171,269){\raisebox{-1.2pt}{\makebox(0,0){$\Diamond$}}}
\put(1191,259){\raisebox{-1.2pt}{\makebox(0,0){$\Diamond$}}}
\put(632,209){\makebox(0,0){$+$}}
\put(669,214){\makebox(0,0){$+$}}
\put(705,221){\makebox(0,0){$+$}}
\put(741,230){\makebox(0,0){$+$}}
\put(778,242){\makebox(0,0){$+$}}
\put(814,259){\makebox(0,0){$+$}}
\put(851,285){\makebox(0,0){$+$}}
\put(887,329){\makebox(0,0){$+$}}
\put(924,412){\makebox(0,0){$+$}}
\put(960,582){\makebox(0,0){$+$}}
\put(997,750){\makebox(0,0){$+$}}
\put(1033,584){\makebox(0,0){$+$}}
\put(1069,413){\makebox(0,0){$+$}}
\put(1106,329){\makebox(0,0){$+$}}
\put(1142,286){\makebox(0,0){$+$}}
\put(1179,259){\makebox(0,0){$+$}}
\put(1215,242){\makebox(0,0){$+$}}
\put(1252,230){\makebox(0,0){$+$}}
\put(1288,221){\makebox(0,0){$+$}}
\put(1324,214){\makebox(0,0){$+$}}
\put(1361,209){\makebox(0,0){$+$}}
\put(312,184){\raisebox{-1.2pt}{\makebox(0,0){$\Box$}}}
\put(381,186){\raisebox{-1.2pt}{\makebox(0,0){$\Box$}}}
\put(449,190){\raisebox{-1.2pt}{\makebox(0,0){$\Box$}}}
\put(518,195){\raisebox{-1.2pt}{\makebox(0,0){$\Box$}}}
\put(586,201){\raisebox{-1.2pt}{\makebox(0,0){$\Box$}}}
\put(654,210){\raisebox{-1.2pt}{\makebox(0,0){$\Box$}}}
\put(723,224){\raisebox{-1.2pt}{\makebox(0,0){$\Box$}}}
\put(791,248){\raisebox{-1.2pt}{\makebox(0,0){$\Box$}}}
\put(860,299){\raisebox{-1.2pt}{\makebox(0,0){$\Box$}}}
\put(928,443){\raisebox{-1.2pt}{\makebox(0,0){$\Box$}}}
\put(997,739){\raisebox{-1.2pt}{\makebox(0,0){$\Box$}}}
\end{picture}
\caption{Fit of a raw data from a Fig. {\protect\ref{su2raw}}
 for a system with 
$SU(2)$ mixing between the levels: 
$\xi_M/M_l$ versus $\epsilon M_{l}^{1/\nu}$ with $\nu =1.1$.}
\label{nu11}
\end{figure}

\begin{figure}
%
\setlength{\unitlength}{0.240900pt}
\ifx\plotpoint\undefined\newsavebox{\plotpoint}\fi
\sbox{\plotpoint}{\rule[-0.200pt]{0.400pt}{0.400pt}}%
\begin{picture}(1500,900)(0,0)
\font\gnuplot=cmr10 at 10pt
\gnuplot
\sbox{\plotpoint}{\rule[-0.200pt]{0.400pt}{0.400pt}}%
\put(220.0,113.0){\rule[-0.200pt]{292.934pt}{0.400pt}}
\put(220.0,113.0){\rule[-0.200pt]{4.818pt}{0.400pt}}
\put(198,113){\makebox(0,0)[r]{0}}
\put(1416.0,113.0){\rule[-0.200pt]{4.818pt}{0.400pt}}
\put(220.0,287.0){\rule[-0.200pt]{4.818pt}{0.400pt}}
\put(198,287){\makebox(0,0)[r]{0.5}}
\put(1416.0,287.0){\rule[-0.200pt]{4.818pt}{0.400pt}}
\put(220.0,460.0){\rule[-0.200pt]{4.818pt}{0.400pt}}
\put(198,460){\makebox(0,0)[r]{1}}
\put(1416.0,460.0){\rule[-0.200pt]{4.818pt}{0.400pt}}
\put(220.0,634.0){\rule[-0.200pt]{4.818pt}{0.400pt}}
\put(198,634){\makebox(0,0)[r]{1.5}}
\put(1416.0,634.0){\rule[-0.200pt]{4.818pt}{0.400pt}}
\put(220.0,808.0){\rule[-0.200pt]{4.818pt}{0.400pt}}
\put(198,808){\makebox(0,0)[r]{2}}
\put(1416.0,808.0){\rule[-0.200pt]{4.818pt}{0.400pt}}
\put(307.0,113.0){\rule[-0.200pt]{0.400pt}{4.818pt}}
\put(307,68){\makebox(0,0){-0.6}}
\put(307.0,857.0){\rule[-0.200pt]{0.400pt}{4.818pt}}
\put(481.0,113.0){\rule[-0.200pt]{0.400pt}{4.818pt}}
\put(481,68){\makebox(0,0){-0.5}}
\put(481.0,857.0){\rule[-0.200pt]{0.400pt}{4.818pt}}
\put(654.0,113.0){\rule[-0.200pt]{0.400pt}{4.818pt}}
\put(654,68){\makebox(0,0){-0.4}}
\put(654.0,857.0){\rule[-0.200pt]{0.400pt}{4.818pt}}
\put(828.0,113.0){\rule[-0.200pt]{0.400pt}{4.818pt}}
\put(828,68){\makebox(0,0){-0.3}}
\put(828.0,857.0){\rule[-0.200pt]{0.400pt}{4.818pt}}
\put(1002.0,113.0){\rule[-0.200pt]{0.400pt}{4.818pt}}
\put(1002,68){\makebox(0,0){-0.2}}
\put(1002.0,857.0){\rule[-0.200pt]{0.400pt}{4.818pt}}
\put(1175.0,113.0){\rule[-0.200pt]{0.400pt}{4.818pt}}
\put(1175,68){\makebox(0,0){-0.1}}
\put(1175.0,857.0){\rule[-0.200pt]{0.400pt}{4.818pt}}
\put(1349.0,113.0){\rule[-0.200pt]{0.400pt}{4.818pt}}
\put(1349,68){\makebox(0,0){0}}
\put(1349.0,857.0){\rule[-0.200pt]{0.400pt}{4.818pt}}
\put(220.0,113.0){\rule[-0.200pt]{292.934pt}{0.400pt}}
\put(1436.0,113.0){\rule[-0.200pt]{0.400pt}{184.048pt}}
\put(220.0,877.0){\rule[-0.200pt]{292.934pt}{0.400pt}}
\put(45,495){\makebox(0,0){$\xi_M/M_l$}}
\put(828,23){\makebox(0,0){$\epsilon$}}
\put(220.0,113.0){\rule[-0.200pt]{0.400pt}{184.048pt}}
\put(1349,848){\raisebox{-.8pt}{\makebox(0,0){$\Diamond$}}}
\put(1262,833){\raisebox{-.8pt}{\makebox(0,0){$\Diamond$}}}
\put(1175,797){\raisebox{-.8pt}{\makebox(0,0){$\Diamond$}}}
\put(1089,749){\raisebox{-.8pt}{\makebox(0,0){$\Diamond$}}}
\put(1002,687){\raisebox{-.8pt}{\makebox(0,0){$\Diamond$}}}
\put(828,560){\raisebox{-.8pt}{\makebox(0,0){$\Diamond$}}}
\put(654,450){\raisebox{-.8pt}{\makebox(0,0){$\Diamond$}}}
\put(481,368){\raisebox{-.8pt}{\makebox(0,0){$\Diamond$}}}
\put(307,309){\raisebox{-.8pt}{\makebox(0,0){$\Diamond$}}}
\put(1349,846){\makebox(0,0){$+$}}
\put(1262,821){\makebox(0,0){$+$}}
\put(1175,790){\makebox(0,0){$+$}}
\put(1089,747){\makebox(0,0){$+$}}
\put(1002,665){\makebox(0,0){$+$}}
\put(828,512){\makebox(0,0){$+$}}
\put(654,388){\makebox(0,0){$+$}}
\put(481,299){\makebox(0,0){$+$}}
\put(307,246){\makebox(0,0){$+$}}
\put(1349,773){\raisebox{-.8pt}{\makebox(0,0){$\Box$}}}
\put(1262,773){\raisebox{-.8pt}{\makebox(0,0){$\Box$}}}
\put(1175,794){\raisebox{-.8pt}{\makebox(0,0){$\Box$}}}
\put(1089,742){\raisebox{-.8pt}{\makebox(0,0){$\Box$}}}
\put(1002,655){\raisebox{-.8pt}{\makebox(0,0){$\Box$}}}
\put(828,463){\raisebox{-.8pt}{\makebox(0,0){$\Box$}}}
\put(654,326){\raisebox{-.8pt}{\makebox(0,0){$\Box$}}}
\put(481,241){\raisebox{-.8pt}{\makebox(0,0){$\Box$}}}
\put(307,195){\raisebox{-.8pt}{\makebox(0,0){$\Box$}}}
\end{picture}

\caption{Renormalized localization length $\xi_M/M_l$ as function of 
energy $\epsilon$ for a system with reduced $U(2)$ mixing 
between the levels ($p=0.3$). $\Diamond$ symbols correspond  
to $M=16$, $+$ to $M=32$ and $\Box$ to $M=64$ system widths. 
A fit to a scaling form yields two extended states at $\epsilon =\pm
0.16$. with $\nu =2.5$}
\label{pdat}
\end{figure}
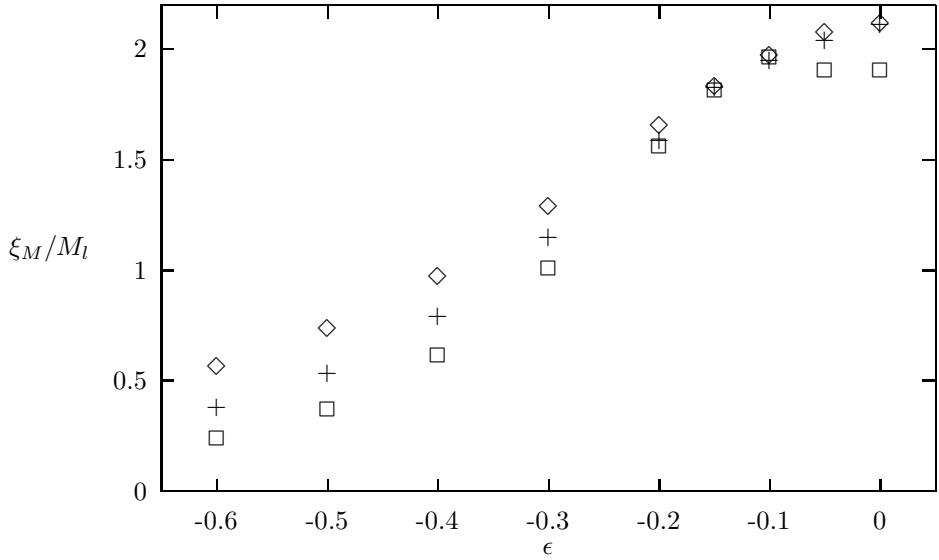

\newpage
\begin{figure}
\input{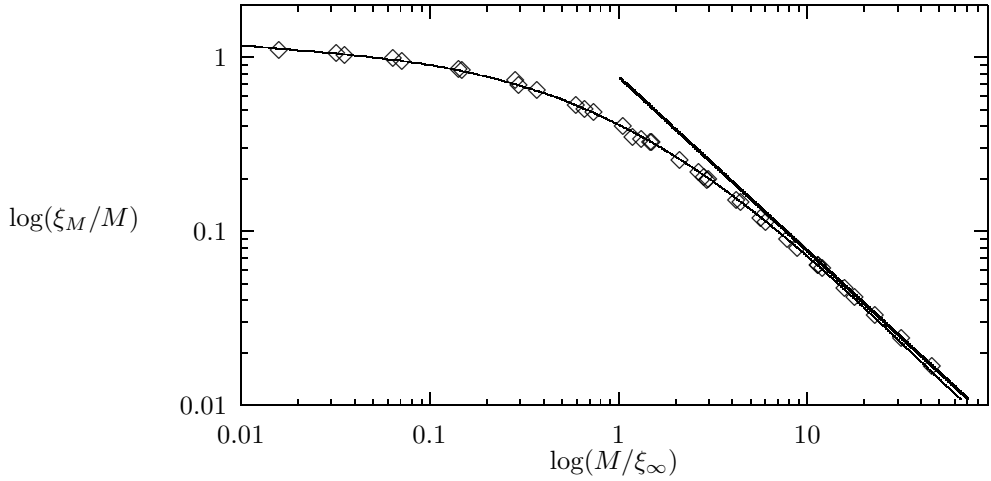}
\caption{A fit of the data for one-channel model: $\xi_M/M$ versus 
$M/\xi$, and a tangential straight line with slope equal to unity.}
\label{beta}
\end{figure}

\begin{table}  
\caption{Symmetry classes of random phases}
\begin{center}
\begin{tabular}{cccc}
Group & link phases $\alpha , \beta$ & \#
 of energies with extended states & $\nu$ \\
\hline
 & & &\\
$SO(2)$ & arbitrary & $1$ & $2.2$\\
 & & &\\
$SU(2)$ & $0$ & $1$ & $1.1$ \\
 & & &\\
$U(2)$ & random & $2$ & $2.5$ \\
\end{tabular}
\end{center}
\end{table}

\begin{references}
\bibitem[*]{}
 Present address: {\it Max-Planck-Institut f\"ur Kernphysik,
Postfach 10 39 80, D-69029, Heidelberg, Germany}.
\bibitem{rqhe} {\it The Quantum Hall Effect}, 
edited by R. E. Prange and S. M. 
Girvin (Springer-Verlag, 1985);
\bibitem{rpruis} A. M. M. Pruisken, {\it Phys. Rev. Lett.}, 
{\bf 61}, 1297(1988);
\bibitem{rtrugman} S. A. Trugman, {\it Phys. Rev. B}, {\bf 27}, 7539 (1983);
\bibitem{rmyl} Milnikov G. V. and Sokolov I. M.,
{\it Pis'ma Zh. Eksp. Teor. Fiz.} {\bf 48}, 494 (1988) [{\it JETP Lett.}, 
{\bf 48}, 536 (1988)];
\bibitem{rAndo} T. Ando, {\it J. Phys. Soc. Japan}, {\bf 53}, 3126 (1984);
\bibitem{rchalk} J. T. Chalker and P. D. Coddington,
        {\it J. Phys. C}, {\bf 21}, 2665 (1988);
\bibitem{rkoch1} S. Koch, R. J. Haug, K. v. Klitzing
  and K. Ploog, {\it Phys. Rev. B}, {\bf 46}, 1596 (1992);
\bibitem{randers} E. Abrahams, P.W. Anderson, D.C. Licciardello  and T.V.
Ramarkrishnan, {\it Phys. Rev. Lett.} {\bf 42}, 673 (1979); 
\bibitem{rnumer} P. A. Lee and D. S. Fisher, {\it Phys. Rev, Lett.} {\bf 47}, 
882 (1981); A. MacKinnon and B. Kramer, {\it Phys. Rev. Lett.} {\bf 47}, 
1546 (1981); A. MacKinnon and B. Kramer in {\it The Application of High 
Magnetic Fields in Semiconductor Physics}, Vol. {\bf 177} of {\it 
Lecture Notes in Physics}, edited by G. Landwehr (Springer, Berlin), p. 74
 (1983);
\bibitem{rkhmel} D.E.Khmelnitskii, 
{\it Phys. Lett} {\bf 106}, 182 (1984); 
Pis'ma Zh. Eksp. Teor. Fiz. {\bf 38}, 454 (1983) [{\it JETP 
Lett.} {\bf 38}, 556 (1983)];
\bibitem{rlaugh} R.B. Laughlin, {\it Phys. Rev. Lett.}, {\bf 52}, 2304 (1984); 
\bibitem{rexp} H.W. Jiang, C.E. Johnson, K.L. Wang and S.T. Hannahs, 
{\it Phys. Rev. Lett.} {\bf 71}, 1439 (1993);
 T. Wang, K.P. Clark, G.F. Spencer, A.M. Mack and W.P. Kirk, 
{\it Phys. Rev. Lett.} {\bf 72}, 709 (1994);
 R.J.F. Hughes, J.T. Nicholls, J.E.F. Frost, 
E.H. Linfield, M. Pepper, C.J.B. Ford, D.A. Ritchie, G.A.C. Jones, E. 
Kogan and M. Kaveh, {\it J. Phys. Condens. Matter} {\bf 6}, 4763 (1994); 
 A.A. Shashkin, G.V. Kravchenko and V.T. Dolgopolov, 
Pis'ma Zh. Eksp. Teor. Fiz. {\bf 52}, 215 (1993)
[{\it JETP 
Lett.} {\bf 58}, 220 (1993)];
 I. Glozman, C.E. Johnson and H.W. Jiang, {\it Phys. Rev. Lett.} 
{\bf 74}, 594 (1995);
\bibitem{rlee}  D. K. K. Lee and J. T. Chalker, {\it Phys. Rev. Lett.}
{\bf 72}, 1510 (1994);
\bibitem{rleeko}  D.K.K. Lee, J. T. Chalker and D. Y. K. Ko, {\it Phys. Rev. B}
{\bf 50}, 5272 (1994);
\bibitem{rwang} Z. Wang, D.H. Lee and X.G. Wen, {\it Phys. Rev. Lett.}
{\bf 72}, 2454 (1994);
\bibitem{rhanna} C. B. Hanna, D. P. Arovas, K. Mullen and S. M. Girvin, 
{\it Phys. Rev. B} {\bf 52}, 5221 (1995);
\bibitem{rjap} K. Minakuchi and S. Hikami, {\it Phys. Rev. B} 
{\bf 53}, 10898 (1996);
\bibitem{rour1} V. Kagalovsky, B. Horovitz and Y. Avishai,
{\it Phys. Rev. B} {\bf 52}, 17044 (1995);
\bibitem{rraih} T. V. Shahbazyan and M. E. Raikh, {\it Phys. Rev. Lett.} 
{\bf 75}, 304 (1995);
\bibitem{dobers} M. Dobers, K. v. Klitzing and G. Weimann, {\it
 Phys. Rev. B} {\bf 38}, 5453 (1988);
\bibitem{rhikami} S. Hikami, M. Shirai and F. Wegner, {\it Nucl. Phys. B} {\bf 
408}, 415 (1993);
\bibitem{dkkl} D. K. K. Lee, {\it Phys. Rev. B} {\bf 50}, 14609
 (1994);
\bibitem {rfert} H. A. Fertig and B. I. Halperin, {\it Phys. Rev. B} {\bf 36},
   7969 (1987);
\bibitem{rhuck1} B. Huckestein and B. Kramer, {\it Phys. Rev. Lett} {\bf 64}, 
1437 {1990};
\bibitem{rour} V. Kagalovsky, B. Horovitz and Y. Avishai, 
{\it Europhys. Lett.} {\bf 31}, 425 (1995);
\bibitem{simplectic} S. N. Evangelou, {\it Phys. Rev. Lett.} {\bf 75}, 
2550 (1995);
\bibitem{Wei} H. P. Wei, S. Y. Lin, D. C. Tsui and A. M. M. Pruisken, 
{\it Phys. Rev. B} {\bf 45}, 3926 (1992);
\bibitem{Lassing} R. Lassnig, {\it Phys. Rev. B} {\bf 31}, 8076 (1985);
\bibitem{Lommer} G. Lommer, F. Malcher and U. R\"{o}ssler, 
{\it Superlattices and Microstructures} {\bf 2}, 273 (1986);
\bibitem{rred} H. E. Stanley, {\it J. Phys. A: Math. Gen.} {\bf 10},
L211 (1977);
\bibitem{rrred} R. Pike and H. E. Stanley, {\it J. Phys. A: Math. Gen.} 
{\bf 14}, L169 (1981);
\bibitem{rred1} A. Coniglio, {\it J. Phys. A: Math. Gen.} {\bf 15},
3829 (1982);
\bibitem{rred2} D. C. Hong and H. E. Stanley, {\it J. Phys. A: Math. Gen.} 
{\bf 16}, L475 (1983);  
\bibitem{rlamb} C. J. Lambert and G. D. Hughes, {\it Phys. Rev. Lett.} 
{\bf 66}, 1074 (1991);
\bibitem{rchalkpr} J. T. Chalker, private communication;
\bibitem{rlevit}  A. Entelis and S. Levit, {\it Phys. Rev. Lett.} {\bf
69}, 3001 (1992); 
\bibitem{rdyk} A. M. Dykhne, Zh. Eksp. Teor. Fiz. {\bf 41}, 1324 (1961)  
[{\it JETP} {\bf 14}, 941 (1962)];
\bibitem{rmood} J. Moody, A. Shapere and F. Wilczek, in {\it Geometric 
Phases in Physics}, edited by A. Shapere and F. Wilczek (World 
Scientific, Singapore, 1989), p. 160.
\end{references}
\end{document}